\documentclass[12pt]{article}
\usepackage{amsmath,amssymb,amsfonts,amsthm,bbm}
\usepackage{epic,eepic,epsfig,longtable}
\usepackage{multirow,verbatim}
\usepackage{array}
\usepackage{graphicx}
\usepackage{floatrow}
\usepackage{subcaption}
\usepackage{paralist}
\usepackage{latexsym}
\usepackage{comment}
\usepackage{epsfig}
\usepackage{setspace}
\usepackage{rotating}
\usepackage{CJK}
\usepackage{color}
\RequirePackage[colorlinks,citecolor=blue,urlcolor=blue]{hyperref}
\usepackage{mathtools}
\DeclarePairedDelimiter{\ceil}{\lceil}{\rceil}
\usepackage[nomarkers, tablesfirst, nolists]{endfloat}
\usepackage{caption}
\usepackage[authoryear,round]{natbib}
\makeatletter
\def\singlespace{\def\baselinestretch{1}\@normalsize}

\makeatletter
\def\singlespace{\def\baselinestretch{1}\@normalsize}

\numberwithin{equation}{section}

\renewcommand{\hat}{\widehat}

\renewcommand{\hat}{\widehat}

    \def\FF{\mathbb{F}}

\newcommand{\bfsym}[1]{\ensuremath{\boldsymbol{#1}}}

\def\1{\bfsym{1}}	

\def\newpage{\vfill\eject}

\def\today{\ifcase\month\or
  January\or February\or March\or April\or May\or June\or
  July\or August\or September\or October\or November\or December\fi
  \space\number\day, \number\year}

\newdimen\biblioindent    \biblioindent=30pt

 at 8truept

\newcommand{\beq}{\begin{equation}}
  \newcommand{\eeq}{\end{equation}}
\newcommand{\beqn}{\begin{eqnarray}}
  \newcommand{\eeqn}{\end{eqnarray}}
\newcommand{\beqnn}{\begin{eqnarray*}}
  \newcommand{\eeqnn}{\end{eqnarray*}}

\allowdisplaybreaks
\usepackage{verbatim}
\def\tilde{\widetilde}

\def\FF{\mathcal{F}}
\def\[{\left [}  \def\]{\right ]} \def\({\left (}  \def\){\right )}
 \def\endpf{$\blacksquare$}
\def\hat{\widehat}

\newtheorem{assumption}{Assumption}
\newtheorem{theorem}{Theorem}
\newtheorem{lemma}{Lemma}

\newtheorem{proposition}{Proposition}
\theoremstyle{definition}
\newtheorem{definition}{Definition}

\newtheorem{remark}{Remark}

\title{Volatility Analysis with Realized GARCH-It\^{o} Models}
\author{Xinyu Song$^1$, 
Donggyu Kim$^2$, 
Huiling Yuan$^3$, 
Xiangyu Cui$^1$,\\
Zhiping Lu$^4$,
Yong Zhou$^{4}$, 
Yazhen Wang$^{4\&5}$\footnote{Corresponding author: Yazhen Wang. Address: 1175 Medical Science Center, 1300 University Avenue ,
Madison, WI 53706. Phone: 6082626399. Fax: 6082620032. E-mail: yzwang@stat.wisc.edu.} \\
\\$^1$ %School of Statistics and Management, \\
Shanghai University of Finance and Economics \\
$^2$ %College of Business, \\
Korea Advanced Institute of Science and Technology (KAIST) \\
$^3$ %School of Statistics and Information, \\
City University of Hong Kong \\
$^4$ %Institute of Statistics and Interdisciplinary Sciences and School of Statistics, \\
East China Normal University \\
$^5$ %Department of Statistics, \\
University of Wisconsin-Madison \\
}

\begin{document}
\maketitle

\begin{abstract}
This paper introduces a unified approach for modeling high-frequency financial data that can accommodate both the continuous-time jump-diffusion and discrete-time realized GARCH model by embedding the discrete realized GARCH structure in the continuous instantaneous volatility process.
The key feature of the proposed model is that the corresponding conditional daily integrated volatility adopts an autoregressive structure, where both integrated volatility and jump variation serve as innovations. 
We name it as the realized GARCH-It\^{o} model.
Given the autoregressive structure in the conditional daily integrated volatility, we propose a quasi-likelihood function for parameter estimation and establish its asymptotic properties.
To improve the parameter estimation, we propose a joint quasi-likelihood function that is built on the marriage of daily integrated volatility estimated by high-frequency data and nonparametric volatility estimator obtained from option data.
We conduct a simulation study to check the finite sample performance of the proposed methodologies and an empirical study with the S\&P500 stock index and option data.
\end{abstract}



%%%%%%%%
%%%%%%%%
%%%%%%%%

%----------------------------------------------------------------------------------------
%	SECTION 1 (Introduction)
%----------------------------------------------------------------------------------------
\newpage 

\renewcommand{\baselinestretch}{1.66}
\baselineskip=22pt

\section{Introduction} 
\label{SEC-1}

In modern financial markets, volatility measures the degree of dispersion for assets and plays a crucial role in portfolio allocation, performance evaluation, and risk management.
Low-frequency and high-frequency stock data are widely adopted to model the dynamic evolution of daily volatilities.
Option data provide one more natural source for the more precise forecast of volatilities and have been investigated thoroughly since the seminal work of \citet{black1973pricing}. 
In traditional volatility analysis, researchers employ discrete parametric econometric models and low-frequency data. 
Examples include the generalized autoregressive conditional heteroskedasticity (GARCH) models \citep{bollerslev1986generalized, engle1982autoregressive} which adopt squared daily log returns as innovations in the conditional volatilities.
However, when the volatility changes rapidly to a new level, it is often difficult to catch up with the new level immediately using only the daily log returns as the innovations \citep{andersen2003modeling}.
On the other hand, high-frequency financial data that refer to intra-daily observations such as tick-by-tick stock prices became available thanks to advances in information technology.
Major challenges in estimating volatilities with high-frequency data are the market microstructure noises and price jumps. 
Without the presence of price jumps, \citet{zhang2005tale} proposed two-time scale realized volatility (TSRV) which is a consistent estimator for daily variation while \citet{zhang2006efficient} further improved the TSRV to multi-scale realized volatility (MSRV) so that it can achieve the optimal convergence rate.
Other forms of estimators that can achieve the optimal convergence rate only in the presence of market microstructure noises are kernel realized volatility (KRV) \citep{barndorff2008designing},  quasi-maximum likelihood estimator (QMLE) \citep{ait2010high, xiu2010quasi},  pre-averaging realized volatility (PRV) \citep{jacod2009microstructure}, and robust pre-averaging realized volatility \citep{fan2018robust}.
Empirical studies support the existence of price jumps, and decomposition of daily variation into its continuous and jump components can improve volatility forecasts \citep{ait2012testing,   andersen2007roughing, barndorff2006econometrics, corsi2010threshold}.
For example, \citet{mancini2004estimation} studied a threshold method for jump-detection and presented the order of an optimal threshold, and \citet{davies2018data} further examined a data-driven type threshold method. 
Also \citet{fan2007multi} and \citet{zhang2016jump} employed wavelet method to identify the jumps given noisy high-frequency data. 
We refer to the estimators of daily variation based on high-frequency data as the realized volatility estimators.
Such estimators are more informative compared to simple squared daily log returns as the innovations, which may help to catch up with rapid changes in the volatility process better.

Efforts made for volatility estimation usually employ low- and high-frequency data independently.
However, the inter-correlation between low- and high-frequency data gathered at the two different time scales cannot be ignored as low-frequency data present high-frequency data in an aggregated form.
There are several attempts to bridge the gap between the two types of data. 
For example, multiple studies proposed new GARCH type models, which include realized volatilities as innovations in the conditional volatilities \citep{engle2006multiple, shephard2010realising, hansen2012realized}.
On the other hand,  \citet{wang2002asymptotic} showed that the standard GARCH model and its diffusion limit are nonequivalent asymptotically, which discredits the direct application of statistical inferences derived for the GARCH model to its diffusion limit. 
Thus, \citet{kim2016unified} introduced the unified GARCH-It\^o model by embedding the standard GARCH volatility structure in the instantaneous volatilities of an It\^o diffusion process. 
The unified GARCH-It\^o model is a continuous-time process at the high-frequency timescale and when restricted to the low-frequency timescale, retains the standard GARCH structure.

In this paper, we expand the unified GARCH-It\^o model \citep{kim2016unified} so that features of financial data at both frequencies can be better captured as follows. 
First, price jumps that are well-documented in empirical studies are allowed, and we incorporate squared price jumps into the volatility dynamics by a structure similar to the ones introduced in the COGARCH model \citep{kluppelberg2004continuous} and the jump-driven volatility model \citep{todorov2011econometric}.
%First, price jumps that are well-documented in empirical studies are allowed, and we incorporate the squared price jumps in the volatility dynamics such as COGARCH model  \citep{kluppelberg2004continuous} and jump-driven volatility model \citep{todorov2011econometric}.
Second, we embed the realized GARCH volatility structure \citep{hansen2012realized} in the instantaneous volatilities of a jump-diffusion process, which employs the more informative high-frequency data-based innovations.
Third, the well-known intra-day U-shape volatility pattern is accounted for \citep{admati1988theory, andersen1997intraday, andersen2019time, hong2000trading}. 
We name the proposed model as the realized GARCH-It\^o model. 
The key feature of the proposed model is that its conditional volatility has integrated volatility and jump variation as innovations.
Based on the structure of the conditional volatility process, we propose a quasi-likelihood function for estimating model parameters.
Specifically, the quasi-likelihood function that is usually adopted in the standard GARCH type models is employed, and the realized volatility estimators are used as the proxy for conditional volatilities.
We call the proposed estimator the quasi-maximum likelihood estimator based on high-frequency data and low-frequency structure (QMLE-HL). 
The proposed model and this estimating approach are constructed purely based on stock data.
We as well harness option data to improve the model parameter estimation. 
In specific, \citet{todorov2019nonparametric} developed nonparametric volatility estimator based on a portfolio of short-dated option contracts given a general setting where jumps are present. 
As stated in \citet{todorov2019nonparametric}, the estimator can be viewed as the option counterpart of high-frequency data-based volatility estimators. 
To incorporate the option-based nonparametric volatility estimator, we construct a joint quasi-likelihood function. 
We call the proposed estimator the quasi-maximum likelihood estimator based on high-frequency data, low-frequency structure and additional option data (QMLE-HLO).
Both the QMLE-HL and the QMLE-HLO present good consistency and asymptotic properties.
In numerical analysis, we further demonstrate that the joint estimation method QMLE-HLO performs better in estimation and prediction than the QMLE-HL.

This paper is organized as follows.
Section \ref{SEC-2} introduces the realized GARCH-It\^o model.
We demonstrate its connection with the realized GARCH model and discuss its advantages comparing to the unified GARCH-It\^o model. 
Section \ref{SEC-3} introduces quasi-likelihood estimation methods and investigates their asymptotic behaviors.
Section \ref{SEC-4} conducts a simulation study to check the finite sample performance for the proposed estimators.
Section \ref{SEC-5} carries out an empirical analysis with S\&P500 stock and option data to demonstrate the advantage of the proposed model in volatility analysis.
We collect all the proofs in the Appendix. 

%%%%%%%%%%%%
%%%%%%%%%%%%
%%%%%%%%%%%%

%----------------------------------------------------------------------------------------
%	SECTION 2 (Realized GARCH-Ito)
%----------------------------------------------------------------------------------------
\setcounter{equation}{0}
\section{Realized GARCH-It\^{o} model} \label{SEC-2}

The realized GARCH-It\^{o} model is an innovated jump-diffusion process that can incorporate high-frequency based volatility model \citep{shephard2010realising} and realized GARCH model \citep{hansen2012realized} structures. 
Let $\mathbb{R}_+=[0, \infty)$ and $\mathbb{N}$ be the set of all non-negative integers.
Our proposed model is formulated as follows.

\begin{definition}
	\label{def:UGARCH}
	Log stock price $X_t$, $t \in \mathbb{R}_+$, obeys a realized GARCH-It\^{o} model if it satisfies
		\begin{equation}
		\label{UGARCH Price}
		dX_t = \mu_t dt + \sigma_t (\theta) dB_t + L_t  d \Lambda_t,
		\end{equation}	
		\begin{eqnarray} \label{UGARCH Vol}
		\sigma^2_t(\theta) &=& \sigma^2_{\ceil{t-1}}(\theta)  + \gamma (t-\ceil{t-1})^2 \left \{ \omega_1 +   \sigma^2_{\ceil{t-1}}(\theta) \right \}  - (t-\ceil{t-1}) \left \{ \omega_2 +   \sigma^2_{\ceil{t-1}}(\theta)   \right \}\cr
		&&  + \alpha   \int_{\ceil{t-1}}^t \sigma^2_s(\theta) ds +\beta  \int_{\ceil{t-1}}^t L_s ^2d \Lambda_s     + \nu \left( \ceil{t-1}+1-t \right) Z^2_t,
		\end{eqnarray}
	where $\ceil{t-1}$ denotes the ceiling of $t-1$, $Z_t = \int_{\ceil{t-1}}^t dW_t$, $B_t$ and $W_t$ are standard Brownian motions with respect to filtration $\FF_t$ with $dW_t dB_t=\rho dt$ a.s., 
	$\mu_t$ is a predictable process that is known as the drift, 
	and $\sigma_t(\theta)$ is the volatility process that is adapted to $\FF_t$.
	For the jump part,  $\Lambda_t$ is the standard Poisson process with constant intensity $\lambda$ and $L_t$ denotes the i.i.d. jump sizes which are independent of the Poisson and continuous diffusion processes.
\end{definition}

\begin{remark}
The i.i.d. assumption on jump sizes can be rewritten as 
	\begin{equation}
	\label{eq:Jump}
	L_t^2=\omega_L+M_t,
	\end{equation}
where $M_t$'s are i.i.d. random variables with mean zero and variance $\zeta^2$, $\omega_L + M_t$ is restricted to be positive. 
For instance, if the jump sizes $L_t$'s obey the Normal distribution with mean $\delta$ and variance  $\eta$, then the corresponding $\omega_L$ takes value $\delta^2 + \eta$ while $M_t$ has mean zero and variance $4 \delta^2 \eta + 2 \eta^2$. 
\end{remark}

The instantaneous volatility $\sigma^2_t(\theta)$ in \eqref{UGARCH Vol} is defined at all times for $t \in \mathbb{R}_+$ and also retains some U-shape pattern within the intra-day.
Specifically, when considering the deterministic process part of the instantaneous volatility, it is convex with respect to time $t$ and for an appropriate parameter, it has the smallest value in the middle section of the day. 
This U-shape instantaneous volatility pattern is often observed in empirical data and supported by financial market \citep{admati1988theory, andersen1997intraday, andersen2019time, hong2000trading}. 
Moreover, random fluctuations are accounted for in the instantaneous volatility process. 
We note that when the process is restricted to integer times, it employs the realized GARCH model type structure \citep{hansen2012realized} with an additional jump innovation term as follows: 
	\begin{equation}
	\label{sig-GARCH}
	\sigma^2_n(\theta)  = \omega  + \gamma  \sigma^2_{n-1}(\theta)  + \alpha  \int_{n-1}^n \sigma^2_s(\theta) ds    + \beta   \int_{n-1}^n L_s ^2d \Lambda_s   , 
	\end{equation}
where $\omega=  \gamma \omega_1  -\omega_2 $ and $n \in \mathbb{N}$.	
Therefore, the instantaneous volatility process is affected by both the integrated volatilities and the jump variations of the stock price process.
In comparison to the unified GARCH-It\^o model \citep{kim2016unified}, the realized GARCH-It\^o model considers price jumps, accounts for intra-day U-shape volatility pattern, and adopts a richer volatility dynamics with random fluctuations.

For statistical inferences, we study the integrated volatilities obtained from the realized GARCH-It\^{o} model over consecutive integers, that is, $\int_{n-1}^n \sigma^2_t(\theta) dt$.
\begin{proposition}
	\label{Lemma : integrated vol}
	Iterative relationship exists in integrated volatilities for the realized GARCH-It\^{o} model defined in Definition \ref{def:UGARCH} and when condition \eqref{eq:Jump} is met.
	\begin{enumerate}
		\item[(a)] For $0<\alpha<1$ and $ n \in \mathbb{N}$,  the realized GARCH-It\^{o} model implies that
			\begin{equation}
			\label{equation-lemma-vol2}
			\int_{n-1}^n \sigma^2_t(\theta)dt = h_n(\theta)+D_n \quad a.s.,
			\end{equation}
		where
			\begin{equation}
			\label{GARCH term}
			h_n(\theta) = \omega^g + \gamma h_{n-1} (\theta) + \alpha^g \int_{n-2}^{n-1} \sigma^2_s(\theta) ds + \beta ^g \int_{n-2} ^{n-1} L_t^2 d \Lambda_t,
			\end{equation}
			\begin{eqnarray}
			\label{GARCH-parameter}
			&& \omega^g = \gamma (\rho_1 -\varrho_2 + 2 \varrho_3) \omega_1 - ( \varrho_1 -  \gamma \varrho_2 +2\gamma \varrho_3 ) \omega_2+ (1-\gamma) \{  (\varrho_{2}-2\varrho_{3})\nu+ \varrho_2 \beta  \lambda \omega_L \}, \cr
			&&  \alpha^g=\left( \rho_1 -\rho_2 +2\gamma \varrho_3 \right) \alpha  , \quad  \beta^g=  \left( \rho_1 -\rho_2 +2\gamma \varrho_3 \right)  \beta,  \quad  \theta = \left(\omega^g, \alpha^g, \beta^g, \gamma \right), \cr
			&&  \rho_1 = \alpha^{-1} (e^{\alpha}-1), \quad \rho_2 = \alpha^{-2} (e^{\alpha}-1-\alpha),  \quad \rho_3=\alpha^{-3} (e^{\alpha}-1-\alpha-\frac{\alpha^2}{2}),
			\end{eqnarray}
			and
			\begin{eqnarray*}
			&&D_n = D_n^c +D_n^J, \cr
			&& D_n^c= 2 \nu \alpha^{-2} \int_{n-1}^n \left\lbrace \alpha (n-t-\alpha^{-1}) e ^{\alpha(n-t)} + 1 \right\rbrace Z_t dZ_t, \cr
			&& D_n^J = \beta  \alpha^{-1} \left \{   \int_{n-1}^{n}  \(e ^{\alpha (n-t)} -1 \) M_t d \Lambda_t  + \omega_L \int_{n-1}^{n}  \(e ^{\alpha (n-t)} -1 \) (d \Lambda_t -\lambda dt )  \right \}
			\end{eqnarray*}
		are all martingale differences.
		
		\item[(b)] For $0<\alpha<1$ and $ n \in \mathbb{N}$,
			\begin{equation}
			\label{equation-cor-GARCH}
			E \left[ \int_{n-1}^n \sigma^2_t (\theta) dt \Bigg| \mathcal{F}_{n-1} \right] = h_n(\theta) \quad a.s.,
			\end{equation}
		where $h_n(\theta)$ is defined in (\ref{GARCH term}).
		
		\item[(c)] For $0< \alpha^g + \gamma <1$ and $n \in \mathbb{N}$,
			\begin{equation}
			\label{GARCH expectation}
			E[h_{n}(\theta)]= \frac{\omega^{g} + \beta^g \lambda \omega_L}{1-\alpha ^g-\gamma}, 
			\quad 
			E[\sigma_{n}^{2}]= \frac{(\omega+ \beta \lambda \omega_L) (1-\alpha^g-\gamma)+\alpha (\omega^{g}+ \beta^g \lambda \omega_L)   }{(1-\alpha ^g-\gamma)(1-\gamma)},
			\end{equation}
		where $\omega^{g}$, $\alpha ^g$ and $\beta^g$ are defined in  \eqref{GARCH-parameter}.
		\end{enumerate}
\end{proposition}

Proposition \ref{Lemma : integrated vol} (a) indicates that the daily integrated volatility can be decomposed into the realized GARCH volatility $h_n(\theta)$ and the martingale difference $D_n$, where the GARCH volatility $h_n(\theta)$ can be further explained by historical integrated volatilities and jump variations. 
We utilize this model feature to build up parameter estimation methods.
Moreover, this paper uses the integrated volatilities as proxy to develop an estimation procedure for the GARCH parameter $\theta = \left(\omega^g, \alpha^g, \beta^g, \gamma \right)$ in Section \ref{SEC-3}.
This is because without the spot volatility estimation, we cannot distinguish the interceptor parameters $\omega_1$, $\omega_2$, and $\nu$.

%%%%%%%%%%%%
%%%%%%%%%%%%
%%%%%%%%%%%%

%----------------------------------------------------------------------------------------
%	SECTION 3 (Parameter Estimation)
%----------------------------------------------------------------------------------------
\setcounter{equation}{0}
\section{Parameter estimation} \label{SEC-3}

In this section, we first discuss the model set-up and review nonparametric estimation methods for the integrated volatility in the presence of market microstructure noises given the jump-diffusion process.
With the well-performing realized volatility and jump variation estimators, we construct quasi-maximum likelihood estimation procedures and investigate their asymptotic behaviors. 

%----------------------------------------------------------------------------------------
%	SECTION 3.1 (Model Set-Up)
%----------------------------------------------------------------------------------------
\subsection{The model set-up and  realized volatility estimators} \label{SEC-3.1}

Let $n$ be the total number of low-frequency observations and $m_i$ be the total number of high-frequency observations during the $i$th low-frequency period, for example, the $i$th day.
We further denote $m=\sum_{i=1}^n m_i/n$.
The underlying log price process is assumed to obey the realized GARCH-It\^{o} model as described in Definition \ref{def:UGARCH}. 
The low-frequency data are the true log prices at integer times, $X_i, i=0,1, \ldots, n$.
The high-frequency data are observations between integer times and are contaminated by market microstructure noises.
Major sources for the market microstructure noises are bid-ask bounce, discreteness of price change, and infrequent trading that only play a role in high-frequency trading \citep{ait2008high}. 
We let $t_{i,j}$ be the high-frequency observed time points during the $i$th low-frequency period such that $i-1=t_{i,0} < t_{i,1} < \cdots < t_{i,m_i} = t_{i+1,0} = i$.
In this regard, we take the well-agreed assumption in high-frequency literature such that 
	\begin{equation} \label{DataModel}
	Y_{t_{i,j}} = X_{t_{i,j}} + \epsilon_{t_{i,j}},
	\end{equation}
where $\epsilon_{t_{i,j}}$'s are market microstructure noises that are some stationary random variables with $E(\epsilon_{t_{i,j}})=0$. %that are independent from the price, the volatility and the jump processes. 
%From a statistical perspective, we assume that $\epsilon_{t_{i,j}}$ are i.i.d. random variables with $E(\epsilon_{t_{i,j}})=0$. 
Moreover, we note that the effect of the drift term $\mu_t$ on high-frequency data based volatility estimators is negligible asymptotically, so we take $\mu_t=0$ to highlight on modeling the volatility and jump processes.

Without the presence of price jumps, researchers have constructed nonparametric realized volatility estimators that take advantage of sub-sampling and local-averaging techniques to remove the effect of market microstructure noises so that the integrated volatility can be estimated consistently and efficiently. 
Such estimators include the multi-scale realized volatility estimator \citep{zhang2006efficient, zhang2011estimating}, the pre-averaging realized volatility estimator \citep{christensen2010pre, jacod2009microstructure}, and the kernel realized volatility estimator \citep{barndorff2008designing}. 
To identify the jump locations given noisy high-frequency data, \citet{fan2007multi} and \citet{zhang2016jump} proposed wavelet methods to detect jumps and applied the MSRV method to jump-adjusted data. 
They demonstrated that the estimator of jump variation has the convergence rate of $m^{-1/4}$, which further helps the estimator of integrated volatility to achieve the optimal convergence rate of $m^{-1/4}$. 
In this paper, we let $JV_i$ to be the estimator of jump variation for the $i$th day and $RV_i$ to be the corresponding estimator of daily integrated volatility that is robust to microstructure noises and price jumps, where both estimators can achieve the convergence rate $m^{-1/4}$. 

%----------------------------------------------------------------------------------------
%	SECTION 3.3 QMLE-HL
%----------------------------------------------------------------------------------------
\subsection{Quasi-maximum likelihood estimation based on high-frequency data and low-frequency structure} \label{SEC-3.3}

\subsubsection{Estimation procedure}

Recall that the integrated volatility over the $i$th period can be decomposed into the realized GARCH volatility $h_i(\theta)$ and martingale difference $D_i$ as described in Proposition \ref{Lemma : integrated vol} (a).
We harness this information for making inferences on the true parameter $\theta_0 =(\omega^g_0, \alpha^g_0, \beta_0^g, \gamma_0)$. 
Specifically, using the likelihood of the standard GARCH model and the low-frequency structure of the realized GARCH-It\^o model, we define the following quasi-likelihood function
	\begin{equation} \label{q-likelihood1} 
	L_{n,m}^{GH}(\theta) =  - \sum\limits_{i=1}^n\left[ \log (h_i(\theta)) + \frac{RV_i}{h_i(\theta)} \right].
	\end{equation}
Under some technical conditions, the impact of the martingale difference term $D_i$ is negligible in the asymptotic sense.
Therefore, the realized volatility estimators $RV_i$'s based on data from \eqref{DataModel} 
can be considered as the observed value for $h_i(\theta)$'s and are employed as the proxy.
To harness the proposed quasi-likelihood function \eqref{q-likelihood1}, we first need to evaluate the realized GARCH term $h_i(\theta)$.
Recall the iterative relationship in the realized GARCH term $h_i(\theta)$ as described in Proposition \ref{Lemma : integrated vol} (a):
	\begin{equation*}
	\begin{split}
	h_i(\theta) = & \omega^g + \gamma h_{i-1}(\theta) + \alpha^g \int_{i-2}^{i-1} \sigma^2_t (\theta) dt  +\beta ^g \int_{i-2}^{i-1} L_t^2 d \Lambda_t \\
	=& \sum\limits_{l=1}^{i-1} \gamma^{l-1} \left\lbrace  \omega^g + \alpha^g \int_{i-l-1}^{i-l}  \sigma^2_t (\theta) dt  +\beta ^g \int_{i-l-1}^{i-l} L_t^2 d \Lambda_t \right\rbrace + \gamma^{i-1} h_1(\theta), \quad i=2, \ldots, n.
	\end{split}
	\end{equation*}
The initial $h_1(\theta)$ is selected to be $E[h_1(\theta)]$ that is given in Proposition \ref{Lemma : integrated vol} (c). Specifically, we take
$$h_1(\theta)=\frac{\omega^g + \beta^g \lambda \omega_L}{1-\alpha^g-\gamma}.$$
The true integrated volatilities and jump variations are not observed so that we adopt their estimators $RV_i$ and $JV_i$, respectively. 
Specifically, let 
	\begin{equation} \label{h-estimate}
	\hat{h}_i(\theta) = \sum\limits_{l=1}^{i-1} \gamma^{l-1} \left\lbrace  \omega^g + \alpha^g RV_{i-l}  + \beta^g JV_{i-l} \right\rbrace + \gamma^{i-1} h_1(\theta), \quad i=2, \ldots,n.
	\end{equation}
With the realized GARCH volatility estimator $\hat{h}_i(\theta)$ in \eqref{h-estimate}, the quasi-likelihood function \eqref{q-likelihood1} is updated to the following:
	\begin{equation} \label{q-likelihood2}
	\hat{L}_{n,m}^{GH}(\theta) =  - \sum\limits_{i=1}^n\left[ \log (\hat{h}_i(\theta)) + \frac{RV_i}{\hat{h}_i(\theta)} \right].
	\end{equation}
We estimate the true parameter $\theta_0$ by maximizing the quasi-likelihood function $\hat{L}_{n,m}^{GH}(\theta) $ in \eqref{q-likelihood2}, 
	\begin{equation} \label{theta-GH}
	\hat{\theta}^{GH} = \underset{\theta \in \Theta }{\mathrm{argmax}} \mbox{ } \hat{L}_{n,m}^{GH} (\theta),
	\end{equation}
and call the maximizer $\hat{\theta}^{GH}$ in \eqref{theta-GH} the quasi-maximum likelihood estimator based on high-frequency data and low-frequency structure combined (QMLE-HL).

%%%%%%%%%
%%%%%%%%%
%%%%%%%%%

\subsubsection{Asymptotic theory}

This section establishes the consistency and asymptotic distribution for the proposed estimator $\hat{\theta}^{GH}$.
We first define some notations.
For any given random variable $X$ and $p \geq 1$, define $\| X \|_{L_p} = \left\lbrace E[|X|^p] \right\rbrace^{1/p}$.
For a matrix $A = \left(A_{i,j} \right)_{1 \leq i \leq k', 1 \leq j \leq k}$, let $\|A\|_{max} = \mbox{max}_{i,j} | A_{i,j} |$.
Let $C$'s be positive generic constants whose values are free of $\theta$, $n$, and $m_i$, and may change from occurrence to occurrence. 
To investigate the asymptotic behaviors of proposed estimation method, we require the following technical assumptions.

\begin{assumption} \label{assumption : GH}
	~
	\begin{enumerate}
		\item[(a)] Let
		\begin{equation*}
		\Theta = \lbrace  (\omega^g, \alpha^g, \beta^g, \gamma): \omega_l^g < \omega^g < \omega_u^g, \alpha_l^g < \alpha^g < \alpha_u^g, \beta_l^g <\beta^g < \beta_u^g, \gamma_l < \gamma < \gamma_u,  \alpha^g + \gamma <1 \rbrace, 
		\end{equation*}
		where $\omega_l^g, \omega_u^g, \alpha_l^g, \alpha_u^g, \beta_l^g, \beta_u^g, \gamma_l, \gamma_u$ are known positive constants.
		
		\item[(b)] We have $\underset{t \in \mathbb{R_+}}{\max} \mbox{ } E \left\lbrace \sigma^4_t(\theta_0) \right\rbrace < \infty $ and $E (\epsilon_{t_{i,j}}^4 ) < \infty$.
		
		\item[(c)] There exist some fixed constants $C_1$ and $C_2$ such that $C_1 m \leq m_i \leq C_2 m$, and $\sup_{1 \leq j \leq m_i} | t_{i,j} - t_{i, j-1} | = O(m^{-1})$ and $n^2m^{-1} \rightarrow 0$ as $m,n \rightarrow \infty$ .
		
		\item[(d)] One of the following conditions is satisfied.
		\begin{itemize}
			\item[(d1)] There exists a positive constant $\delta$ such that $E \left[ \left( \frac{R_{i}^{2}}{h_{i}(\theta_0)} \right) ^{2+\delta} \right] \leq C$ for any $i \in \mathbb{N}$, where $R_i=\int_{i-1}^i \sigma_t (\theta_0) dB_t$.
			
			\item[(d2)] $\frac{E[R_i^4 | \mathbf{\mathcal{F}}_{i-1}]}{h^2_i(\theta_0)} \leq C$ a.s. for any $i \in \mathbb{N}$.
		\end{itemize}
		
		\item[(e)] $\sup\limits_{i\in \mathbb{N}} \left\|RV_{i}-\int_{i-1}^{i}\sigma_{s}^{2} (\theta_0)ds\right\|_{L_{2}}\leq C   m^{-1/4}$ and 
		$\sup\limits_{i\in \mathbb{N}} \left\|JV_{i}-\int_{i-1}^{i}L_{s}^{2} d \Lambda_s \right\|_{L_{2}}\leq C   m^{-1/4}$.
		
		\item[(f)] For any $i\in \mathbb{N}$, $E\left[RV_{i}|\mathcal{F}_{i-1}\right]\leq C \, E \left[\int_{i-1}^{i}\sigma_{s}^{2}ds | \mathcal{F}_{i-1} \right]+C$ a.s.
		
		\item[(g)] $\left(D_i, \int_{i-1}^i \sigma^2_{t}(\theta_0)dt, R^2_i \right)$ is a stationary ergodic process.
	\end{enumerate}
\end{assumption}

\begin{remark}
The parameters of interests are related to volatilities (the 2nd moment), thus, to study their asymptotic behaviors, we require some finite 4th moment conditions such as Assumption \ref{assumption : GH} (b) and (d).
Therefore, these conditions are not restrictive at all.
Assumption \ref{assumption : GH} (c) is a well-known key condition in high-frequency data based volatility analysis.
Under the finite 4th moment condition, \citet{kim2016asymptotic} showed that the realized volatility estimators satisfy Assumption \ref{assumption : GH} (e).
Finally, the stationary ergodic condition Assumption \ref{assumption : GH} (g) is used to obtain asymptotic normality for the QMLE-HL. 
\end{remark}

The following theorems establish the convergence rate and asymptotic normality for the QMLE-HL $\hat{\theta}^{GH}$ defined in \eqref{theta-GH}. 

\begin{theorem}
\label{Thm-GH-rate}
Under Assumption \ref{assumption : GH} (a)-(f) (except for $n^2 m^{-1} \rightarrow 0$ in Assumption \ref{assumption : GH} (c)), we have
	\begin{equation*}
	\left \| \hat{\theta}^{GH} - \theta_0 \right \|_{max} = O_p \left( m^{-1/4} + n^{-1/2} \right).
	\end{equation*}
\end{theorem}

\begin{theorem}
\label{Thm-GH-normality}
Under Assumption \ref{assumption : GH}, we have as $m,n \rightarrow \infty$,
	\begin{equation*}
	\sqrt{n} \left( \hat{\theta}^{GH} -\theta_0 \right) \overset{d}{\rightarrow} N \left(0, B^{-1} A^{GH} B^{-1} \right),
	\end{equation*}
where
	\begin{eqnarray*}
	A^{GH} &=&  E \Bigg[  \Bigg \{ \alpha_0^{-4} \nu_0^2  \int _{0}^{1} \left \{  \alpha_0(1-t  -\alpha_0^{-1}) e^{\alpha_0(1-t)}  +1 \right \} ^2 t dt \cr
	&& \qquad  +\frac{ \lambda_0 \beta_0^2  }{4\alpha_0^{2}}   \int_{0}^{1}  \(e ^{\alpha_0 (1-t)} -1 \)^2 (M_t ^2 +\omega_{L0}^2) d  t  \Bigg \}    \frac{\partial h_1(\theta)}{\partial \theta  } \frac{\partial h_1(\theta )}{\partial \theta  ^{T}} \Bigg|_{\theta=\theta_0} h_1^{-4}(\theta_0)\Bigg]
	\end{eqnarray*}
and 
	\begin{equation*}
	B =\frac{1}{2} E \left[ \frac{\partial h_1(\theta )}{\partial \theta } \frac{\partial h_1(\theta)}{\partial \theta^T } \Bigg| _{\theta=\theta_0}  h_1^{-2}(\theta_0 ) \right]. 
	\end{equation*}
\end{theorem}

\begin{remark}
Theorem \ref{Thm-GH-rate} shows that the convergence rate of $\hat{\theta}^{GH}$ is $m^{-1/4} + n^{-1/2}$.
The rate $n^{-1/2}$ is coming from the usual parametric convergence rate based on the low-frequency structure while the rate $m^{-1/4}$ is due to the high-frequency volatility and jump variation estimations and is known as the optimal convergence rate for estimating integrated volatilities with the presence of market microstructure noises and price jumps. 
Theorem \ref{Thm-GH-normality} provides the asymptotic normal distribution for $\hat{\theta}^{GH}$.
When deriving the asymptotic normality, the condition $n^2 m^{-1} \rightarrow 0$ in Assumption \ref{assumption : GH} (c) is imposed so that the high-frequency estimation errors of order $m^{-1/4}$ are negligible in comparison with the low-frequency estimation errors of order $n^{-1/2}$.
% in Theorem \ref{Thm-GH-normality}. 
When the condition $n^2 m^{-1} \rightarrow 0$ is not satisfied, the asymptotic normality may depend on $m^{1/4}  (RV_i - \int_{i-1}^{i}\sigma_{s}^{2} (\theta_0)ds)$, which is the quantity related to high-frequency estimation. 
For example, if $m^{1/4}  (RV_i - \int_{i-1}^{i}\sigma_{s}^{2} (\theta_0)ds)$ is some martingale difference sequence, we can relax the condition $n^2 m^{-1} \rightarrow 0$ to $nm^{-1} \rightarrow 0$.
We also note that if the true stock prices are observed (i.e., without the microstructure noises), we only need the typical condition $n m^{-1} \rightarrow 0$ instead of $n^2 m^{-1} \rightarrow 0$ to obtain the asymptotic normality (see \citet{todorov2009estimation}). 
\end{remark}

\begin{remark}
We note that when replacing $m^{-1/4}$ in Assumption \ref{assumption : GH} (e) by $m^{-\xi}$ for some positive constant $\xi  \in (0, 1/4]$, the convergence rate in Theorem \ref{Thm-GH-rate} will change to $m^{-\xi} + n^{-1/2}$. 
On the other hand, the condition $n^2 m^{-1} \rightarrow 0$ in Assumption \ref{assumption : GH} (c) will be relaxed to $n^2 m^{-4 \xi} \rightarrow 0$ for deriving the asymptotic normality in Theorem \ref{Thm-GH-normality}. 
%Specifically, there exist two types of estimation errors: the errors coming from estimating the realized volatility and jump variation with high-frequency data and have the rate of $m^{-1/4}$; the errors coming from estimating model parameters with the quasi-likelihood function that is constructed on the low-frequency structure and have the rate of $n^{-1/2}$. 
%The condition $n^2 m^{-1} \rightarrow 0$ in Assumption \ref{assumption : GH} (c) handles the estimation errors coming from estimating the realized volatility and jump variation, which is required only to derive the asymptotic normality in Theorem \ref{Thm-GH-normality}. 
%Furthermore, when replacing $m^{-1/4}$ in Assumption \ref{assumption : GH} (e) by $m^{-\xi}$ for some positive constant $0<\xi \leq 1/4$, we will obtain convergence rate $m^{-\xi} + n^{-1/2}$ in Theorem \ref{Thm-GH-rate}.   
%On the other hand, when the condition $n^2 m^{-1} \rightarrow 0$ is not satisfied, the asymptotic normality may depend on the structure of  $m^{1/4}  (RV_i - \int_{i-1}^{i}\sigma_{s}^{2} (\theta_0)ds)$. 
%For example, if $m^{1/4}  (RV_i - \int_{i-1}^{i}\sigma_{s}^{2} (\theta_0)ds)$ is some martingale difference sequence, we can relax the condition $n^2 m^{-1} \rightarrow 0$ to $nm^{-1} \rightarrow 0$.
%which is usually true in the high-frequency setup. 
\end{remark}

%----------------------------------------------------------------------------------------
%	SECTION 3.4 QMLE-HLO
%----------------------------------------------------------------------------------------

\subsection{Quasi-maximum likelihood estimation based on based on high-frequency data, low-frequency structure, and additional option data} \label{SEC-3.4}

\subsubsection{Estimation procedure}

In this section, we discuss how to incorporate additional option data information in parameter estimation.
The famous Black-Scholes model indicates that option prices are determined by several factors such as time to expiration, strike price, underline asset price, and its volatility, and so one can deduce the volatility from option data. 
For example, the VIX presents the stock market's general expectation of volatility. 
However, we usually find that the VIX is different from the historical nonparametric realized volatility.
This may be because of the jumps in stock prices and the wedge between the risk-neutral and statistical probabilities.
Recently, \citet{todorov2019nonparametric} proposed a nonparametric volatility estimator based on a portfolio of noisy short-dated option contracts with different strike prices.
This estimator is robust to price jumps and does not require any assumption on the wedge between risk-neutral and statistical probabilities. 
Specifically, let $T$ be the time to expiration for an option contract, $k_\ell$ be the $\ell$th log strike price, where $k_1 < k_2< \cdots <k_N$ and $\Delta_\ell=k_\ell-k_{\ell-1}$ for $\ell=2, \ldots, N$. 
Let $\kappa_T (k_\ell)$ be the true option price given expiration $T$ and log-strike $k_\ell$.
Due to observation errors in empirical derivatives pricing, the observed option price $\hat{\kappa}_T (k_\ell)$ obeys
	\begin{equation*} 
	\hat{\kappa}_T (k_\ell) = \kappa_T (k_\ell) + \varepsilon _\ell,
	\end{equation*}
where the noises $\varepsilon _\ell$'s are random variables with mean zero and satisfy the technical conditions in \citet{todorov2019nonparametric}.
Given this set-up, \cite{todorov2019nonparametric} proposed the following nonparametric volatility estimator
	\begin{equation*}
	NV_i=\frac{-2}{Tu}  \mathcal{R} \( \log \( \hat{f}_i ( u ) \wedge T\) \),
	\end{equation*}
where $$\hat{f}_i (u) = 1-(u^2+ \sqrt{-1} u) \sum_{\ell=2}^N e^{(\sqrt{-1} u-1) k_{\ell-1} -\sqrt{-1} u X_i} \hat{\kappa}_T (k_{\ell-1}) \Delta_\ell,$$
$\mathcal{R}(A)$ is the real part of a complex number $A$, and $u$ is a tuning parameter.

Under some technical conditions, as $T$ goes to zero, this nonparametric volatility estimator $NV_i$ converges to the true spot volatility $\sigma^2_i(\theta_0)$  \citep{todorov2019nonparametric}.
However, option contracts from traditional data sources such as the OptionMetrics are often quoted at the market open or close on each trading day so that the minimum choice of $T$ is $1$ business day. 
In this sense, $NV_i$ may contain integrated volatility for the remaining period from time $i$. 
Also \citet{todorov2019nonparametric} showed that the estimates $NV_i$'s hold a close relationship with the jump-robust realized type volatility estimates $RV_i$'s in his empirical study. 
Based on his results, we assume that the nonparametric volatility estimator $NV_{i-1}$ and the conditional daily integrated volatility $h_{i}(\theta)$ have the following linear relationship: 
	\begin{equation}
	\label{eq: option non vol}
	NV_{i-1}=b + a h_{i} (\theta) + e_i, \quad  i=1, \ldots, n,
	\end{equation}
where $b$ and $a$ are the intercept and slope coefficients, respectively. 
Moreover, $e_i$'s are martingale differences with mean zero and variance $\sigma_e^2$, and they are independent of the price process and the microstructure component.

Let $\varphi=(\omega^g, \alpha^g, \beta^g, \gamma, a, b)$ and $\phi=(\omega^g, \alpha^g, \beta^g, \gamma, a, b, \sigma^2_e)$.
Note that $\theta$ corresponds to the first four coordinates of $\varphi$ and $\phi$. 
We generalize \eqref{q-likelihood2} to propose the following joint quasi-likelihood function based on high-frequency and option data for estimating the true parameter $\phi_0=(\omega^g_0, \alpha^g_0, \beta^g_0, \gamma_0, a_0, b_0, \sigma^2_{e0})$
	\begin{equation} \label{q-likelihood3}
	\hat{L}^{GHO}_{n,m}(\phi) = - \sum_{i=1}^n \left[ \log (\hat{h}_i(\theta)) +\frac{RV_i}{ \hat{h}_i(\theta)} \right] - \sum_{i=1}^{n} \left[ \log ( \sigma_e^2 ) + \frac{(NV _{i-1} - b - a \hat{h}_{i}(\theta))^2}{\sigma_e^2} \right].
	\end{equation}
We maximize $\hat{L}^{GHO}_{n,m}(\phi)$ in \eqref{q-likelihood3} to obtain  parameter estimators, that is, 
	\begin{equation} \label{theta-GHO}
	\hat{\phi}^{GHO}= \underset{\phi \in \Phi}{\mbox{argmax }} \hat{L}_{n,m}^{GHO}(\phi), \;\;\; \hat{\theta}^{GHO}=\mbox{the first four coordinates of }  \hat{\phi}^{GHO}, 	
	\end{equation}
where $ \Phi$ is the parameter space of $\phi$.
We call the proposed estimator $\hat{\phi}^{GHO}$ (or $\hat{\theta}^{GHO}$) in \eqref{theta-GHO} 
 the quasi-maximum likelihood estimator based on high-frequency data, low-frequency structure, and additional option data combined (QMLE-HLO).

%\begin{remark}
%This is because practically, T ranges from 1 to 5 business days so that $NV_i$ contains volatility information for the remaining period. 
%Another approach is to connect $NV_i$ with the spot volatility at integer time $i$, that is, $\sigma^2_i(\theta_0)$, from the realized GARCH-It\^o model. 
%In fact, Theorems 1 and 3 in  \citet{todorov2019nonparametric} show that as $T$ goes to 0, the nonparametric volatility estimator $NV_i$ converges to the spot volatility $\sigma_i^2 (\theta_0)$.
%Similar to $\hat{L}^{GHO}_{n,m}(\phi)$, we could construct a joint quasi-likelihood function by taking advantage of the asymptotic limits provided in \citet{todorov2019nonparametric}. 
%In this case, the convergence rate of the corresponding estimator will depend on the convergence rate of $NV_i$, which may not be optimal. 
%Specifically, Theorem 1 in \citet{todorov2019nonparametric} shows that $NV_i$ converges to $\sigma_i^2(\theta_0)$ with the convergence rate $T^{\tau}$, where $\tau$ is a positive number that is determined by the range of log-strikes and the choice of $u$. 
%In this case, $T^{\tau}$ plays the dominant role in the convergence rate of the quasi-maximum likelihood estimator, which is rather slow due to the fixed remaining period of $T$ in practice. 
%Also the numerical study shows that the QMLE-HLO method has better performance and thus, we present the QMLE-HLO method in this paper. 
%\end{remark}

%%%%%%%%%%
%%%%%%%%%%
%%%%%%%%%%

\subsubsection{Asymptotic theory}

To establish the asymptotic behaviors of the proposed estimation method, we require the following additional assumptions. 

\begin{assumption} \label{assumption-GHO}
	~
	\begin{enumerate}
		\item[(a)] Let
		\begin{equation*}
				\Phi = \lbrace (\omega^g, \alpha^g, \beta^g, \gamma, a, b, \sigma_{e}^{2}):  (\omega^g, \alpha^g,  \beta^g, \gamma) \in \Theta,    a_l < a < a_u, b_l < b < b_u, \sigma_{e_{l}}^{2} < \sigma_{e}^{2} < \sigma_{e_{u}}^{2} \rbrace, 
		\end{equation*}
		where $a_l, a_u, b_l,  b_u, \sigma_{e_{l}}^{2}, \sigma_{e_{u}}^{2}$ are known positive constants.
		\item [(b)] $\sup_{i \in \mathbb{N}} E \[ e_i^4\] < \infty$. 
		\item[(c)] $\left(D_i, \int_{i-1}^i \sigma^2_{t}(\phi_0)dt, R^2_i,  e_i \right)$ is a stationary ergodic process. 
	\end{enumerate}
\end{assumption}

The following theorems establish the convergence rate and asymptotic normality for the QMLE-HLO $\hat{\phi}^{GHO}$
defined in \eqref{theta-GHO}. 

\begin{theorem}
\label{Thm-GHQ-rate2}
Under Assumption \ref{assumption : GH} (a)--(f) (except for $n^2 m^{-1} \rightarrow 0$ in Assumption \ref{assumption : GH} (c)) and Assumption \ref{assumption-GHO} (a)--(b), we have
	\begin{equation*}
	\left\|\hat{\phi}^{GHO}-\phi_{0}\right\|_{max} = O_{p} \left( n^{-1/2} + m^{-1/4} \right).
	\end{equation*}
\end{theorem}

\begin{theorem}\label{Thm-GHQ-normality2}
Under Assumption \ref{assumption : GH} and Assumption \ref{assumption-GHO}, we have as $m, n\rightarrow \infty$,
	\begin{equation*}
	\sqrt{n}\left(\hat{\phi}^{GHO}-\phi_{0}\right)\xrightarrow{d}N\left(0,\left(B^{GHO}\right)^{-1}A^{GHO}\left(B^{GHO}\right)^{-1}\right),
	\end{equation*}
where
	\begin{equation*}
	A^{GHO}=  \left(
	\begin{tabular}{cc}
	$A^{GH}$ & $\mathbf{0}_{4\times3}$ \\
	$\mathbf{0}_{4\times3}^{T}$ & $\mathbf{0}_{3\times3}$ \\
	\end{tabular}
	\right) + A^{O}, 
	\qquad
	B^{GHO}=\left(
	\begin{tabular}{cc}
	$B^{\varphi}$ & $\mathbf{0}_{6\times1}$ \\
	$\mathbf{0}_{6\times1}^{T}$ & $\frac{1}{2} \sigma^{-4}_{e0}$ \\
	\end{tabular}
	\right),
	\end{equation*}
	\begin{align*}
	A^{O}=& E\[    \left(
	\begin{tabular}{cc}
	$\frac{\partial f_{1}(\varphi)}{\partial \varphi}\frac{\partial f_{1}(\varphi)}{\partial\varphi^{T}}\bigg|_{\varphi=\varphi_{0}}\frac{1}{\sigma_{e0}^2}$ & $\frac{\partial f_{1}(\varphi)}{\partial \varphi} \bigg|_{\varphi=\varphi_{0}}\frac{e_1^3 }{2\sigma_{e0}^6}$ \\
	$\frac{\partial f_{1}(\varphi)}{\partial \varphi ^{T} } \bigg|_{\varphi=\varphi_{0}}\frac{e_1 ^3  }{2\sigma_{e0}^6}$ & $\frac{ (e_1^2 - \sigma_{e0}^2) ^2}{4\sigma_{e0}^8}$ \\
	\end{tabular}
	\right)  \]	,\\
	B^{\varphi}=& \frac{1}{2} E\left[\frac{\partial h_{1}(\theta)}{\partial\varphi}\frac{\partial h_{1}(\theta)}{\partial\varphi^{T}}\bigg|_{\varphi=\varphi_{0}}h_{1}^{-2}(\theta_{0}) + \frac{\partial f_{1}(\varphi)}{\partial\varphi}\frac{\partial f_{1}(\varphi)}{\partial\varphi^{T}}\bigg|_{\varphi=\varphi_{0}}\frac{2}{\sigma_{e0}^2}\right],
	\end{align*}
and $f_i(\varphi)= b+a h_i(\theta)$ for $i=1, \ldots, n$. Here $\mathbf{0}_{i\times j}$ denotes an $i$-by-$j$ matrix of zeros.
\end{theorem}

\begin{remark}
Theorem \ref{Thm-GHQ-rate2} shows that the convergence rate for the QMLE-HLO is the same as the QMLE-HL.
Theorem \ref{Thm-GHQ-normality2} provides the asymptotic normal distribution for the QMLE-HLO.
\end{remark}

%%%%%%%%%%
%%%%%%%%%%
%%%%%%%%%%
 
%----------------------------------------------------------------------------------------
%	SECTION 4 (Simulation Study)
%----------------------------------------------------------------------------------------
\setcounter{equation}{0}
\section{Simulation study} \label{SEC-4}

In this section, we conducted a simulation study to check the finite sample performance of the estimators $\hat{\theta}^{GH}$ and $\hat{\phi}^{GHO}$ given by \eqref{theta-GH} and \eqref{theta-GHO} respectively, 
as well as to investigate the prediction performance of the realized GARCH volatilities $\hat{h}_i(\hat{\theta}^{GH})$ and $\hat{h}_i(\hat{\theta}^{GHO})$, which was also compared with the performance of the GARCH volatilities used in \citet{kim2016unified}. Here $\hat{h}_i( \cdot) $ is defined in \eqref{h-estimate}. 
The true log prices $X_{t_{i,j}}$, $t_{i,j}=i-1+j/m$, $i=1, \ldots, n$, $j=1, \ldots, m$, were generated based on the proposed realized GARCH-It\^{o} model defined in \eqref{UGARCH Price} and \eqref{UGARCH Vol} with the following set of parameters $\omega_1=5.816$, $\omega_2=1.228$, $\alpha=0.765$, $\beta=0.482$, $\nu=0.6$, $\gamma=0.225$, and $\rho=-0.6$. 
For the jump process, we took the intensity $\lambda$ to be 26 and generated $L_t^2$ such that $L_t^2=\omega_L+M_t$, where $\omega_L=0.005$ and $M_t$ follows the normal distribution with mean zero and standard deviation 0.001.
Each jump $L_t$ was further assigned to be either positive or negative randomly. 
The chosen parameters resulted in the following target parameter $\theta=(\omega^g, \alpha^g, \beta^g, \gamma)=(0.0122, 0.717, 0.452, 0.225)$ for modeling the dynamics in conditional integrated volatilities. 
We note that the parameter $\omega^g$ was scaled by 10000 times compared to its empirical counterpart while the rest parameters remained the same.    
Scaling in this simulation study was done in order to avoid the generation of any negative value for the instantaneous volatilities due to the U-shape intra-day pattern. 
Initial values for the simulation were chosen to be $X_0=10$ and $\sigma_{0}^{2}=E(\sigma_{1}^{2})=1.4$.
For the high-frequency data $Y_{t_{i,j}}$'s from \eqref{DataModel}, market microstructure noises were added to simulated log prices $X_{t_{i,j}}$'s between integer times, and the noises were modeled by i.i.d normal random variables with mean $0$ and standard deviation $0.005$.
For the option model described in \eqref{eq: option non vol}, we took $a=0.812$, $b=0.072$, $\sigma_e=0.04$, where the intercept $b$ and standard deviation $\sigma_{e}$ were scaled by roughly 10000 times comparing to their empirical estimates. 
We took $n=125,250,500,1000$ and $m=390, 780, 2340, 23400$.
For each combination of $n$ and $m$, we repeated the simulation procedure for 2000 times.
We followed the procedure as described in \citet{fan2007multi} to detect the jump locations, estimate the jump variations, and compute the jump-adjusted MSRV estimators.
Model parameter estimators were obtained by maximizing the proposed quasi-likelihood functions $\hat{L}^{GH}_{n,m}(\theta)$ and $\hat{L}^{GHO}_{n,m}(\phi)$ defined in \eqref{q-likelihood2} and \eqref{q-likelihood3}, respectively.

Table \ref{Table: jump parameters} reports the mean squared errors (MSEs) for the jump parameters $\omega_L$ and $\lambda$. 
We find that the MSEs decrease as the number of high-frequency observations increases for each $n$, and larger $n$ often helps to locate the jumps and to estimate the parameters $\omega_L$ and $\lambda$ better.
Table \ref{Table: parameters} presents the MSEs for the QMLE-HL and QMLE-HLO.
The proposed estimating procedures present good finite sample performances and support the theoretical results derived in Section \ref{SEC-3}.
For each estimation method, as the number of low-frequency or high-frequency observations increases, the MSEs decrease.
When comparing the two methods, the QMLE-HLO has smaller MSE than the QMLE-HL.
Thus, it is reasonable to conclude that additional option data help to enhance the estimation of model parameters. 
%Thus, we can conjecture that additional option data help to improve the estimation of model parameters. 

\begin{sidewaystable}[h]
%\begin{table}[h]
	\centering
	\begin{tabular}{ccccccccccc}
	\hline 
	\hline 
	&& \multicolumn{9}{c}{\textbf{MSE}} \\
	\cline{3-11}
	&& \multicolumn{4}{c}{$\omega_L$} && \multicolumn{4}{c}{$\lambda$} \\ 
	\cline {3-6} \cline {8-11} 
	$n$ \textbackslash \mbox{ } $m$ && $390$ & $780$ & $2340$ & $23400$ && $390$ & $780$ & $2340$ & $23400$ \\
	\hline 
	$125$ && $5.461 \times 10^{-4}$ & $1.288 \times 10^{-4}$ & $1.317 \times 10^{-5}$ & $4.149 \times 10^{-8}$ &&457.691&329.315&180.923&1.868\\
	$250$ && $5.335 \times 10^{-4}$ & $1.244 \times 10^{-4}$ & $1.231 \times 10^{-5}$ & $3.960 \times 10^{-8}$ &&456.705&327.808&177.606&1.609\\
	$500$ && $5.232 \times 10^{-4}$ & $1.213 \times 10^{-4}$ & $1.190 \times 10^{-5}$ & $3.921 \times 10^{-8}$ &&453.558&327.112&176.499&1.480\\
	$1000$ && $5.182 \times 10^{-4}$ & $1.193 \times 10^{-4}$ & $1.159 \times 10^{-5}$ & $3.859 \times 10^{-8}$ &&450.895&325.991&175.006&1.227\\
	\hline 
	\hline 
	\end{tabular}
\caption{The mean squared errors (MSEs) for the jump process parameters $\omega_L$ and $\lambda$ given $n=125,250,500,1000$ and $m=390,780,2340,23400$.
\label{Table: jump parameters}}
%\end{table}
\end{sidewaystable}

\begin{sidewaystable}[ph!]
	\centering
	\begin{tabular}{ccccccccccccccc}
		\hline
		\hline
		&&& \multicolumn{12}{c}{\textbf{MSE} $\mathbf{\times 10^3}$} \\
		\cline{3-15}
		&&& \multicolumn{4}{c}{\textbf{QMLE-HL}} && \multicolumn{7}{c}{\textbf{QMLE-HLO}} \\
		\cline{3-7} \cline{9-15}
		$n$&$m$ && $\omega^g$ & $\alpha^g$ & $\beta^g$ & $\gamma$ && $\omega^g$ & $\alpha^g$ & $\beta^g$ & $\gamma$ & $a$ & $b$ & $\sigma_{e}$ \\
		\hline 
		\multirow{4}{*}{125} & 390 && 22.514 & 83.956 & 401.751 & 80.986 & & 6.349 & 70.967 & 220.209 & 73.972 & 13.829 & 7.301 & 2.707 \\ 
		& 780 && 13.874 & 59.539 & 263.366 & 64.944 && 1.946 & 52.000 & 64.073 & 55.349 & 5.973 & 5.622 & 1.456 \\
		& 2340 && 12.759 & 27.847 & 154.776 & 32.416 && 1.549 & 21.814 & 55.391 & 27.515 & 3.915 & 5.341 & 0.896 \\ 
		& 23400 && 11.414 & 9.172 & 100.052 & 12.784 & & 1.430 & 2.500 & 36.110 & 2.417 & 1.801 & 2.574 & 0.057 \\ 
		&&&&&&&&&&&&&& \\
		\multirow{4}{*}{250} & 390 && 9.197 & 76.620 & 266.625 & 75.162 && 3.862 & 69.965 & 169.024 & 65.612 & 11.865 & 3.784 & 2.018 \\
		& 780 && 4.645 & 50.045 & 146.061 & 58.639 && 1.106 & 46.422 & 34.844 & 50.483 & 4.224 & 3.189 & 1.426 \\ 
		& 2340 && 3.604 & 20.791 & 73.631 & 25.116 && 0.850 & 19.384 & 27.946 & 20.633 & 2.154 & 2.947 & 0.723 \\ 
		& 23400 && 3.089 & 4.571 & 47.478 & 5.838 && 0.762 & 1.356 & 16.774 & 1.209 & 1.135 & 1.557 & 0.029 \\ 
		&&&&&&&&&&&&&& \\
		\multirow{4}{*}{500} & 390 && 4.552 & 71.620 & 187.886 & 69.883 && 2.633 & 65.360 & 140.817 & 60.363 & 10.524 & 2.300 & 1.895 \\ 
		& 780 && 1.767 & 46.471 & 71.798 & 53.530 && 0.561 & 42.864 & 18.012 & 45.275 & 2.939 & 1.983 & 1.357 \\ 
		& 2340 && 1.232 & 17.835 & 42.019 & 18.183 && 0.421 & 13.502 & 16.107 & 15.762 & 1.288 & 1.873 & 0.597 \\ 
		& 23400 && 1.108 & 2.127 & 24.276 & 2.645 && 0.390 & 0.718 & 8.779 & 0.609 & 0.611 & 0.841 & 0.014 \\ 
		&&&&&&&&&&&&&& \\
		\multirow{4}{*}{1000} &390 && 2.544 & 69.202 & 139.960 & 60.467 && 1.808 & 60.128 & 126.474 & 52.889 & 9.530 & 1.694 & 1.646 \\ 
		& 780 && 0.706 & 44.901 & 34.083 & 44.476 && 0.293 & 38.988 & 8.461 & 36.868 & 1.942 & 1.569 & 1.174 \\ 
		& 2340  && 0.522 & 16.317 & 23.354 & 13.971 && 0.271 & 10.613 & 7.610 & 8.862 & 0.855 & 1.222 & 0.500 \\ 
		& 23400 && 0.454 & 1.087 & 13.779 & 1.301 &  & 0.247 & 0.366 & 4.518 & 0.306 & 0.325 & 0.436 & 0.007 \\ 
		\hline
		\hline
	\end{tabular}
	\caption{The mean squared errors (MSEs) for the QMLE-HL and QMLE-HLO methods on estimating realized GARCH volatility parameters for $n=125,250,500,1000$ and $m=390,780,2340, 23400$.
	\label{Table: parameters}}
\end{sidewaystable}

The major motivation of our model proposal is to predict future volatilities by taking advantage of the imposed autoregressive type of model structure at the low-frequency.
So we examined the finite sample performance of the proposed predictors $\hat{h}_i(\hat{\theta}^{GH})$ and $\hat{h}_i(\hat{\theta}^{GHO})$, where $\hat{\theta}^{GH}$ and $\hat{\theta}^{GHO}$ are defined in \eqref{theta-GH} and \eqref{theta-GHO}, respectively, and $\hat{h}_i (\cdot)$ is given by \eqref{h-estimate}. 
For comparison purpose, we as well investigated the prediction performance of the unified GARCH-It\^o model proposed by \citet{kim2016unified}, and denote the predictor by $\hat{h}_{i0}(\hat{\theta}_0^{GH})$. 
%On the other hand, practically, we often employ some martingale assumption, which uses the integrated volatility from previous period as the predictor.
%We also studied the performance of the jump-adjusted MSRV estimator.
Specifically, we evaluated the mean squared prediction errors (MSPEs)  by
	\begin{equation*}
	\frac{1}{n-h} \sum\limits_{i=h+1}^n \left( \hat{H}_i  - h_i(\theta) \right)^2,
	\end{equation*}
where $\hat{H}_i$ is one of the followings: $\hat{h}_i (\hat{\theta}^{GH})$, $\hat{h}_i (\hat{\theta}^{GHO})$, or $\hat{h}_{i0}(\hat{\theta}_0^{GH})$. As a benchmark, we as well considered the prediction of $h_i(\theta)$ using $RV_{i-1}$. 
We let the initial forecast origin to be $h=n-20$ and expanded the observation window by one low-frequency period at a time. Each time, the model parameters were estimated and the predictors were obtained.

\begin{figure}
\centering
\includegraphics[width=\textwidth]{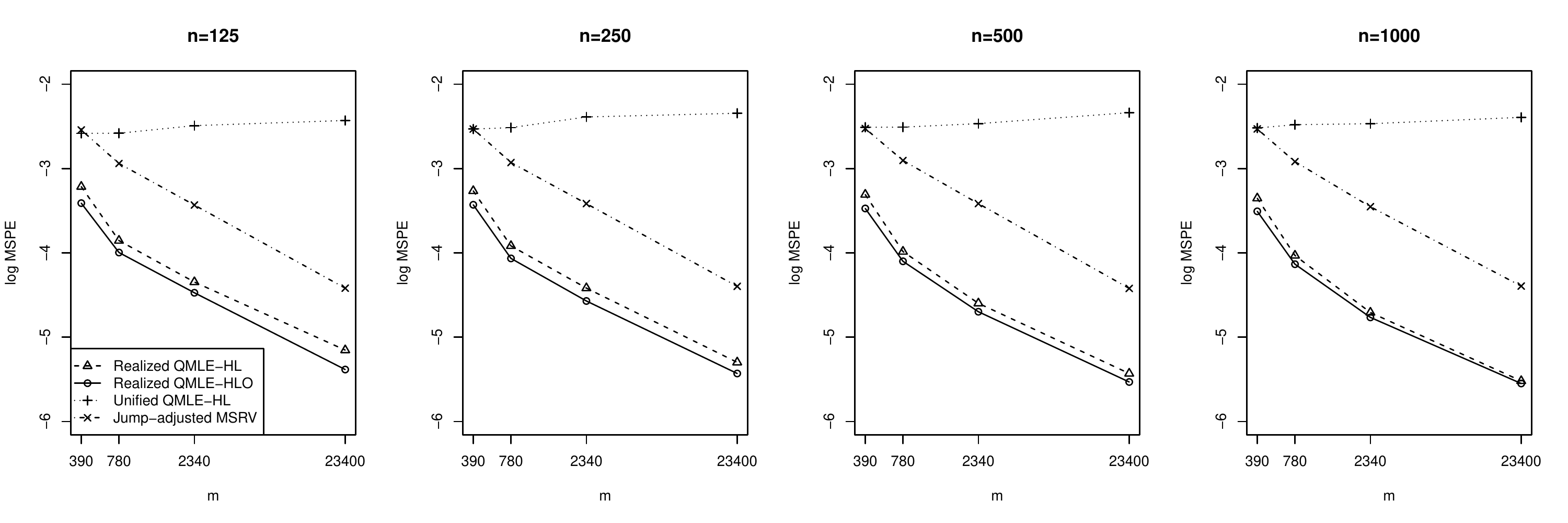}
\caption{The log mean squared prediction errors (MSPEs) of the realized GARCH volatility predictors $h_i(\theta)$, the GARCH volatility predictors $h_{i0}(\theta_0)$ and the benchmark jump-adjusted MSRV predictors $RV_{i-1}$ against $m$ for the different $n$ choices. 
\label{Figure: forecast}}
\end{figure}

Table \ref{Table: GARCH volatilities} summarizes the MSPEs and Figure \ref{Figure: forecast} presents the log MSPEs against the number of high-frequency observations. 
Overall, the MSPE  for the realized GARCH-It\^o approach decreases as the number of low-frequency or high-frequency observations increases. 
Moreover, the QMLE-HLO method presents the best performance regarding the MSPE.
That is, the numerical results indicate that utilizing information contained in an additional data source can improve both the estimation and prediction performance of the proposed methodology.
On the other hand, the unified GARCH-It\^o model is not capable of explaining the rich dynamics in order to predict the conditional integrated volatilities.
This may be because it takes into account neither the realized volatility nor the jump variation as an innovation.
The benchmark method does not perform well because the realized GARCH-It\^o model has rich dynamics that cannot be fully captured by the jump-adjusted MSRV method. 

\begin{sidewaystable}
%\begin{table}[h]
	\centering
	\begin{tabular}{cccccc}
		\hline
		\hline
		&& \multicolumn{4}{c}{\textbf{MSPE} $\mathbf{\times 10^2}$} \\
		\cline{3-6} 
		&& \multicolumn{2}{c}{\textbf{Realized GARCH-It\^o}} & \textbf{Unified GARCH-It\^o} & \textbf{Jump-adjusted} \\
		$n$ & $m$ & \textbf{QMLE-HL} & \textbf{QMLE-HLO} & \textbf{QMLE-HL} & \textbf{MSRV} \\
		\hline
		\multirow{4}{*}{125} & 390 & 4.017 & 3.303 & 7.560 & 7.869 \\ 
		& 780 & 2.119 & 1.839 & 7.570 & 5.287 \\ 
		& 2340 & 1.296 & 1.141 & 8.284 & 3.229 \\ 
		& 23400 & 0.578 & 0.459 & 8.806 & 1.205 \\ 
		&&&&& \\
		\multirow{4}{*}{250} &390 & 3.819 & 3.240 & 7.957 & 7.959 \\ 
		& 780 & 1.990 & 1.715 & 8.088 & 5.346 \\ 
		& 2340 & 1.206 & 1.035 & 9.182 & 3.284 \\ 
		& 23400 & 0.500 & 0.438 & 9.593 & 1.231 \\ 
		&&&&& \\
		\multirow{4}{*}{500} &390 & 3.657 & 3.101 & 8.127 & 8.004 \\ 
		& 780 & 1.860 & 1.657 & 8.138 & 5.478 \\ 
		& 2340 & 1.007 & 0.911 & 8.483 & 3.286 \\ 
		& 23400 & 0.438 & 0.396 & 9.664 & 1.202 \\ 
		&&&&& \\
		\multirow{4}{*}{1000} & 390 & 3.501 & 2.998 & 8.052 & 7.963 \\ 
		& 780 & 1.775 & 1.601 & 8.378 & 5.403 \\ 
		& 2340 & 0.903 & 0.852 & 8.474 & 3.165 \\ 
		& 23400 & 0.401 & 0.389 & 9.141 & 1.235 \\ 
		\hline
		\hline
	\end{tabular}
	\caption{The mean squared prediction errors (MSPEs) of the realized GARCH volatility predictors $h_i(\theta)$ proposed in realized GARCH-It\^o model with the QMLE-HL and the QMLE-HLO methods, the GARCH volatility predictor $h_{i0}(\theta_0)$ proposed in unified GARCH-It\^o model \citep{kim2016unified}, and the benchmark jump-adjusted MSRV method for $n=125,250,500,1000$ and $m=390,780,2340, 23400$. 
	\label{Table: GARCH volatilities}}
%\end{table}
\end{sidewaystable}

%%%%%%%%%%%%
%%%%%%%%%%%%
%%%%%%%%%%%%

%----------------------------------------------------------------------------------------
%	SECTION 5 (Empirical Analysis)
%----------------------------------------------------------------------------------------
\section{Empirical analysis} \label{SEC-5}

\begin{figure}[h]
	\centering
	\includegraphics[scale=0.8]{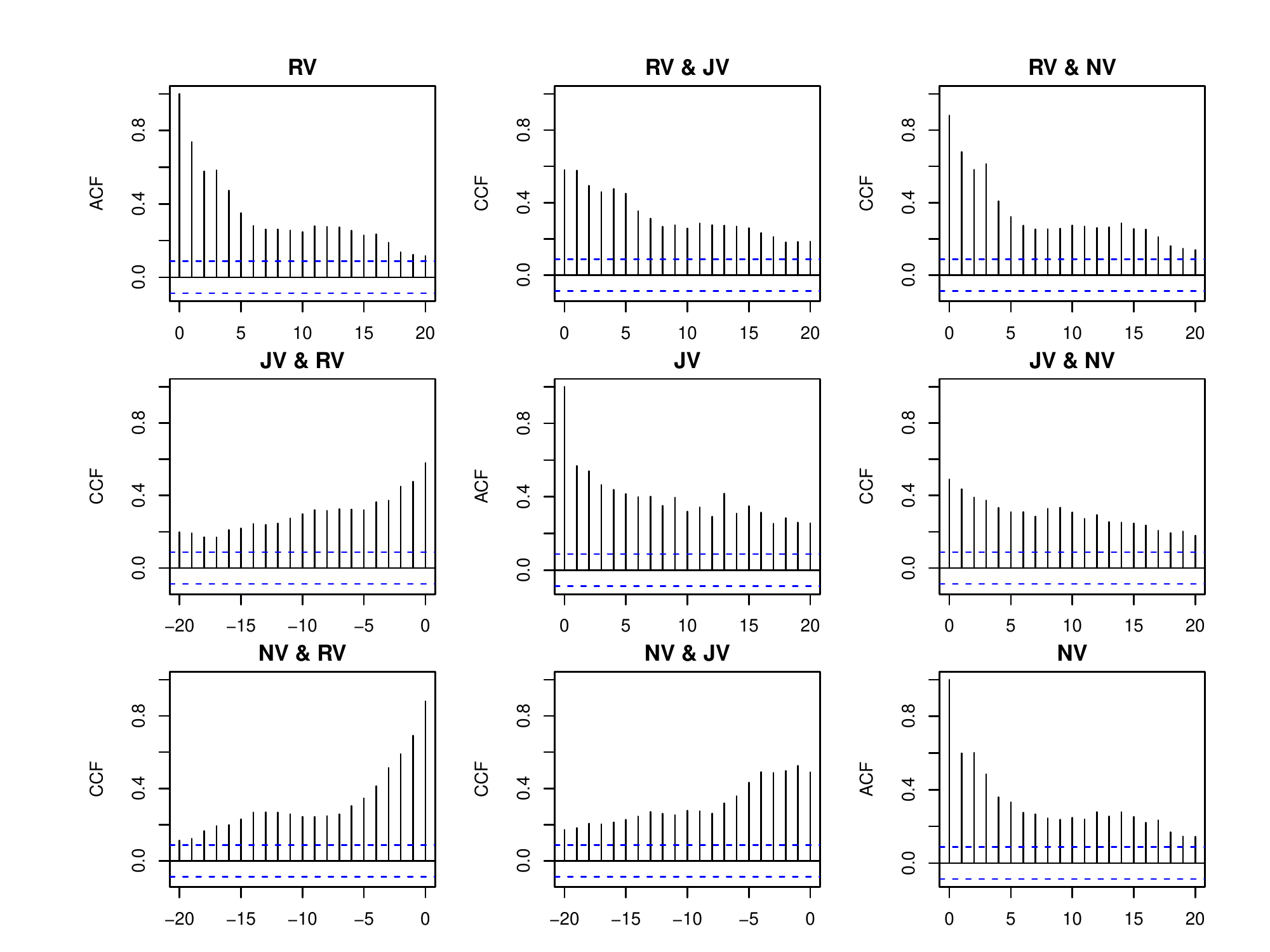}
	\caption{Auto-correlation function (ACF) and cross-correlation function (CCF) plots \citep{brockwell2016introduction} for the time series of the daily jump-adjusted MSRV (RV) estimators, the daily jump variation (JV) estimators and the daily nonparametric volatility (NV) estimators with option data. \label{Figure: ACF}}
	\end{figure}

\begin{figure}[h]
\centering
\includegraphics[width=\textwidth]{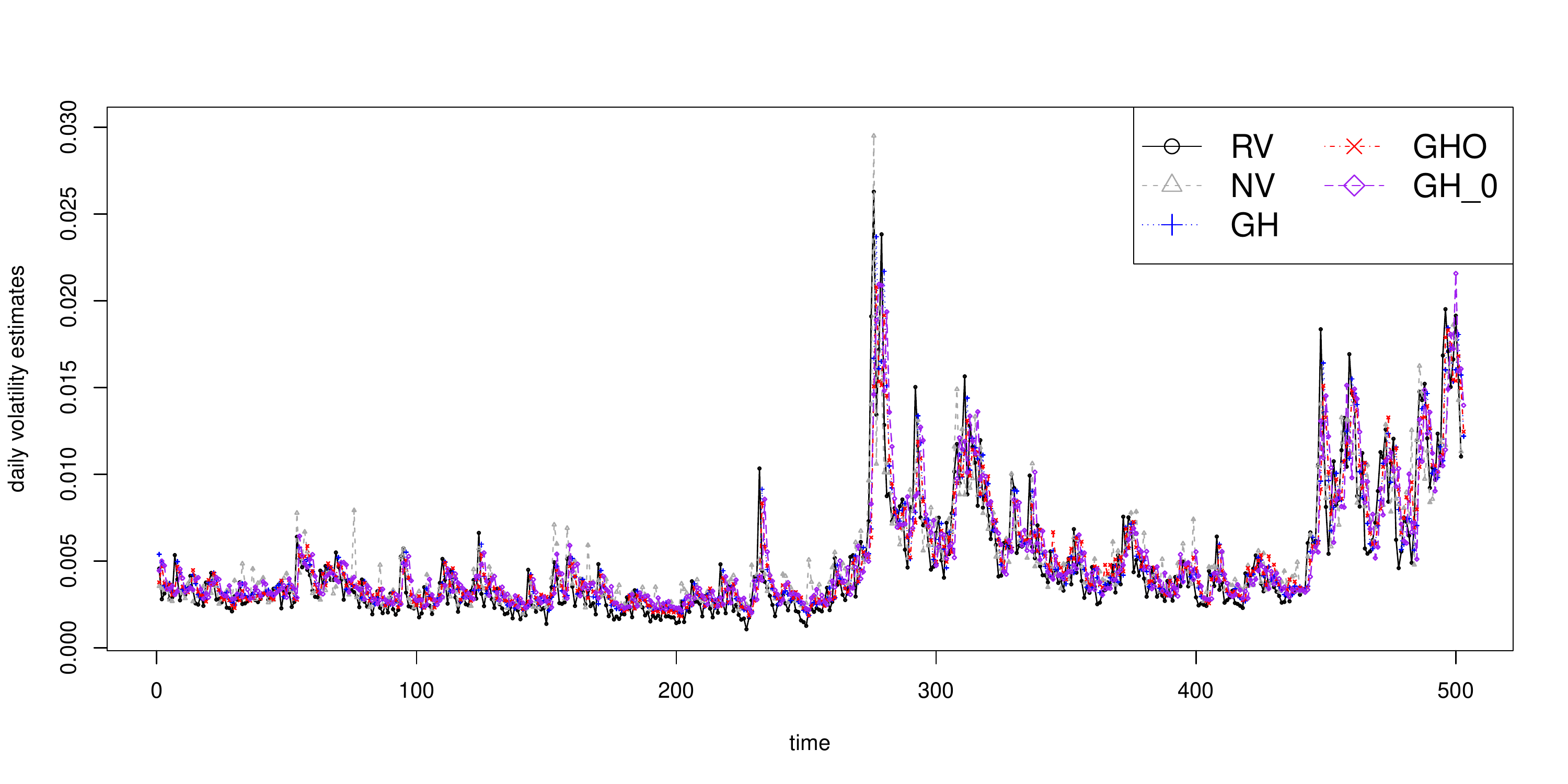}
\vspace*{-30pt}
\caption{Daily volatility estimates with 1) RV: jump-adjusted MSRV estimates $\sqrt{RV_i}$;
2) NV: option-based nonparametric volatility estimates: $\sqrt{NV_i}$; 
3) GH: realized GARCH volatility estimates $\sqrt{\hat{h}_i(\hat{\theta}^{GH})}$ with the QMLE-HL;
4) GHO: realized GARCH volatility estimates $\sqrt{\hat{h}_i(\hat{\theta}^{GHO})}$ with the QMLE-HLO;
5) GH\_0: GARCH volatility estimates $\sqrt{h_{i0}(\hat{\theta}_0^{GH})}$ given the unified GARCH-It\^{o} model.
\label{Figure: Estimated Volatility}}
\end{figure}

In this section, we illustrate the proposed estimation methods with trading data in second for S\&P500 stock index and option data quoted at the market opening on each trading day, where S\&P500 stock index is the underline asset.
The data sets were obtained from the TAQ and the CBOE database, respectively.
We examined the period from January 3rd, 2017 to December 31th, 2018 so that the number of low-frequency periods is $n=502$.
The high-frequency data are available between open and close of the market so that the number of high-frequency observations for a full trading day is $m=23400$. 
We followed the procedure given in \citet{fan2007multi} to detect jumps, as well as to compute the jump variation estimates $JV_i$'s and jump-adjusted MSRV estimates $RV_i$'s.
We estimated the intensity $\lambda$ by the daily averaged number of price jumps, and the parameter $\omega_L$ by the sample median of all squared price jumps because the sample median better described the center of the distribution formed by squared jumps. 
The estimated values are $\hat{\lambda}=25.938$ and $\hat{\omega}_L=3.675 \times 10^{-8}$.
For the option data, we followed the procedure presented in \citet{todorov2019nonparametric} as their empirical study covered a similar period and considered the S\&P500 index as well. 
Specifically, we took the option contracts where the time to expiration ranges from 1 to 2 business days and skipped the contracts that were settled on a holiday. 
The average number of strikes per date was $62.843$ and the values of the tuning parameters were set to be the same as in \citet{todorov2019nonparametric}. 
Denote the option-based nonparametric volatility estimates by $NV_i$'s.
Figure \ref{Figure: ACF} displays the auto- and cross-correlation functions \citep{brockwell2016introduction} for the $RV_i$'s,  $JV_i$'s, and  $NV_i$'s, which provides promising evidence for explaining the rich dynamics with these innovations. 
The QMLE-HL estimates are 
$\hat{\omega}^g=1.224 \times 10^{-6}, \hat{\alpha}^g=0.717, \hat{\beta}^g=0.452$, and $\hat{\gamma}=0.225$, 
and the QMLE-HLO estimates are 
$\hat{\omega}^g=3.450 \times 10^{-7}, \hat{\alpha}^g=0.512,  \hat{\beta}^g=2.375, \hat{\gamma}=0.305, \hat{a}=0.812, \hat{b}=7.198 \times 10^{-6}, \hat{\sigma}_e=4.298 \times 10^{-6}$. 
The parameter $\omega^g$ denotes the intercept term in the realized GARCH volatility dynamics while the parameter $b$ denotes the intercept term in model \eqref{eq: option non vol}.
Their small estimated values reflect the overall level of daily volatilities that can be seen in Figure \ref{Figure: Estimated Volatility}.

Figure \ref{Figure: Estimated Volatility} displays the jump-adjusted MSRV estimates, the option-based nonparametric volatility estimates, the realized GARCH volatility estimates from the QMLE-HL and the QMLE-HLO. 
For comparison purpose, we as well present the GARCH volatilities adopted in the unified GARCH-It\^{o} model \citep{kim2016unified}.
Figure \ref{Figure: Estimated Volatility} shows that the nonparametric jump-adjusted MSRV and the option-based nonparametric volatility estimates are both volatile, 
and the realized GARCH volatility estimates from the QMLE-HL and QMLE-HLO methods can account for these dynamics well. 
Moreover, when comparing with the unified GARCH-It\^{o} estimates, the proposed realized GARCH-It\^{o} estimates are closer to the jump-adjusted MSRV estimates.
This may be because the realized GARCH-It\^{o} model includes realized volatilities and jump variations as innovations while the unified GARCH-It\^{o} model comprises squared daily log returns as innovations.
That is, the proposed structure in the realized GARCH-It\^o model helps to capture the market dynamics promptly. 

To investigate the prediction performance of the proposed methodologies, we employed the MSPE criteria again. 
Denote the forecast origin by $h$. 
To further examine the dependency of split points, we took $h=376, 397, 420, 439, 462, 483$, where each value corresponds to the last trading day of June, July, August, September, October, and November in the year of 2018. 
Since the exact conditional daily integrated volatilities are unknown for empirical data, we used the jump-adjusted MSRV estimates instead and evaluated the following MSPE: 
	\begin{equation*}
	\frac{1}{n-h} \sum\limits_{i=h+1}^{n} \left( \hat{H}_i -  RV_i \right)^2,
	\end{equation*}
where $\hat{H}_i$ is one of the followings: $\hat{h}_i (\hat{\theta}^{GH})$, $\hat{h}_i (\hat{\theta}^{GHO})$, $\hat{h}_{i0}(\hat{\theta}_0^{GH})$, or $RV_{i-1}$, and $\hat{h}_i( \cdot) $ is defined in \eqref{h-estimate}.

Table \ref{empirical: MSPE} summarizes the MSPEs from the realized GARCH-It\^o,  the unified GARCH-It\^o, and the jump-adjusted MSRV estimates. 
Overall, the proposed realized GARCH-It\^{o} estimates outperform the other methods in terms of the MSPE across various split points.
When comparing the realized GARCH-It\^{o} estimates, the QMLE-HLO presents smaller MSPE than the QMLE-HL. 
The empirical results indicate that the realized GARCH-It\^o model holds advantages in predicting future volatilities as it utilizes the autoregressive structure in daily integrated volatilities and emphasizes high-frequency based information by using both realized volatilities and jump variations as innovations. 
Moreover, incorporating option-based nonparametric volatility estimates could help to predict future volatilities. 

\begin{sidewaystable}
%\begin{table}
	\centering
	\begin{tabular}{ccccc}
		\hline
		\hline
		& \multicolumn{4}{c}{\textbf{MSPE} $\mathbf{\times 10^{9}}$} \\
		\cline{2-5}
		 \textbf{Forecast Origin} & \multicolumn{2}{c}{\textbf{Realized GARCH-It\^{o}}} & \multicolumn{1}{c}{ \textbf{Unified GARCH-It\^{o}}} & \textbf{Jump-adjusted} \\
		& \textbf{QMLE-HL} & \textbf{QMLE-HLO} & \textbf{QMLE-HL} & \textbf{MSRV}  \\
		\hline
		$h=376$ & 2.527 & 2.323 & 3.141 & 2.655 \\ 
		$h=397$ & 3.024 & 2.770 & 3.744 & 3.177 \\ 
		$h=420$ & 3.851 & 3.510 & 4.766 & 4.040 \\ 
		$h=439$ & 5.005 & 4.536 & 6.189 & 5.251 \\ 
		$h=462$ & 4.052 & 3.913 & 6.813 & 4.134 \\ 
		$h=483$ & 6.628 & 5.073 & 12.559 & 6.578 \\ 
		\hline  
		\hline
	\end{tabular}
	\caption{The mean squared prediction errors (MSPEs) of the realized GARCH-It\^{o} estimates with the QMLE-HL and the QMLE-HLO, the unified GARCH-It\^{o} estimates with the QMLE-HL, and the jump-adjusted MSRV estimates.
	\label{empirical: MSPE}}
%\end{table}
\end{sidewaystable}

%-----------------------------------------------------------------------------------------------
% Conclusion
%-----------------------------------------------------------------------------------------------

\section{Conclusion} \label{SEC-6}

In this paper, we introduce a novel realized GARCH-It\^{o} model based on a jump-diffusion process which embeds the discrete realized GARCH model structure \citep{hansen2012realized} in its instantaneous volatility process. 
When the model is restricted to the low-frequency period, it employs an autoregressive type structure to explain the co-dynamics in the integrated volatilities and jump variations.  
Model parameters in the realized GARCH-It\^{o} model are estimated by maximizing a quasi-likelihood function.
To improve the statistical performance of the proposed estimating approach and to incorporate additional information from option data, we as well connect the nonparametric volatility estimator proposed by \citet{todorov2019nonparametric} with the conditional integrated volatility from the proposed model. 
A joint quasi-likelihood function is then adopted and we show that this method helps to improve accounting for the market dynamics in the numerical analysis.

We also leave some open issues for future study. 
For example, we may observe some heterogeneous variance in model \eqref{eq: option non vol}. 
One possible approach is to generalize the homogeneous variance in \eqref{eq: option non vol} to heterogeneous variance such as replacing $\sigma_e^2$ by $\sigma_e^2 h_i^{\zeta}(\theta)$, where parameter $\zeta>0$ is used to adjust the level of heteroscedasticity with $\zeta=0$ corresponding to the homogeneous case.
We replace $\sigma_e^2$ by $\sigma_e^2 \hat{h}_i^{\zeta}(\theta)$ in the quasi-likelihood $\hat{L}^{GHO}_{n,m}(\phi)$ given by 
\eqref{q-likelihood3} 
and then estimate $\zeta$ jointly with the other parameters by maximizing $\hat{L}^{GHO}_{n,m}(\phi)$. 
Moreover, it is important to explore further about the optimal approach to combine and model the return and option data for volatility estimation. 
%Moreover, it is important to study the optimal way to combine the return and option data.
%Specifically, we may consider more complex relationships than the linear structure in model \eqref{eq: option non vol} and use estimators other than the $NV_i$'s. 

%\vskip 14pt
%\noindent {\large\bf Supplementary Materials}

%%%%%%%%%%%
%%%%%%%%%%%
%%%%%%%%%%%

\newpage
\vskip 14pt
\appendix
\numberwithin{equation}{section}

\section{Appendix} \label{Appendix}

Let $C>0$ and $0 < \rho < 1$ be generic constants whose values are free of $\theta$, $\phi$, $n$, and $m$ and may change from occurrence to occurrence.

\subsection{Proof of Proposition \ref{Lemma : integrated vol}}

\textbf{Proof of Proposition \ref{Lemma : integrated vol}.}
For $k,n \in \mathbb{N}$, let
	\begin{equation*}
	R(k)\equiv  \int_{n-1}^{n}\frac{(n-t)^{k}}{k!}\sigma_{t}^{2}(\theta)dt.
	\end{equation*}
By the It\^o's Lemma, we have
	\begin{align*}
	R(k)= 
	&\frac{(k+1)\nu}{(k+3)!} +\frac{\gamma  \omega_1 -\omega_2  +  \gamma     \sigma_{n-1}^{2} (\theta)  }{(k+1)!}  +\frac{ \omega_2 - 2 \gamma \omega_1 + (1-2\gamma) \sigma_{n-1}^{2} (\theta)   }{(k+2) k!} \\
	&  +\frac{\gamma \omega_1 + \gamma \sigma_{n-1}^{2} (\theta)  }{(k+3)k!}  +\frac{\beta  \lambda \omega_L }{(k+2)!}  +\alpha R(k+1) \\
	&+\beta  \int_{n-1}^{n}  \frac{(n-t)^{k+1}}{(k+1)!} M_t d \Lambda_t+ \beta \int_{n-1}^n\frac{(n-t)^{k+1}}{(k+1)!} \omega_L (d \Lambda_t - \lambda dt)  \\
	&+2 \nu \int_{n-1}^{n}\left(\frac{(n-t)^{k+2}}{(k+1)!}-\frac{(n-t)^{k+2}}{(k+2)!}\right)Z_{t}dZ_{t} .
  \end{align*}
Then simple algebraic manipulations show  
	\begin{align*}
	&\int_{n-1}^{n} \sigma_{t}(\theta)^{2}dt = R(0) \\
	&=  (\varrho_{2}-2\varrho_{3})\nu+ \varrho_2 \beta  \lambda \omega_L + 2 \varrho_3 \gamma \omega_1 - \varrho_2 \omega_2 + (\varrho_1- \varrho_2 + 2\gamma \varrho_3)  \sigma_{n-1}^{2} (\theta)  +D_n^J+ D_n ^c \quad \text{a.s.}
	\end{align*}
Since
	\begin{equation*}
	\sigma^2_n(\theta) =  \omega  + \gamma  \sigma^2_{n-1}(\theta)  + \alpha  \int_{n-1}^n \sigma^2_s(\theta) ds    + \beta  \int_{n-1}^n L_s ^2d \Lambda_s  , 
	\end{equation*}
we have
	\begin{align*}
	h_{n}(\theta)
	&=   (\varrho_{2}-2\varrho_{3})\nu+ \varrho_2 \beta  \lambda \omega_L + 2 \varrho_3 \gamma \omega_1 - \varrho_2 \omega_2 + (\varrho_1- \varrho_2 + 2\gamma \varrho_3)  \sigma_{n-1}^{2} (\theta)  \\
	&=  (\varrho_{2}-2\varrho_{3})\nu+ \varrho_2 \beta  \lambda \omega_L + 2 \varrho_3 \gamma \omega_1 - \varrho_2 \omega_2   \\
	&\quad  + (\varrho_1- \varrho_2 + 2\gamma \varrho_3)  \left( \omega  + \gamma  \sigma^2_{n-2}(\theta)  + \alpha  \int_{n-2}^{n-1} \sigma^2_s(\theta) ds    + \beta   \int_{n-2}^{n-1} L_s ^2d \Lambda_s   \right)   \\
	&=  \omega^g + \gamma h_{n-1} (\theta) + \alpha^g \int_{n-2}^{n-1} \sigma^2_{s} (\theta) ds + \beta^g  \int_{n-2}^{n-1} L_t^2 d \Lambda_t ,
	\end{align*}
where $\omega^g$, $\alpha^g$ and $\beta^g$ are defined in \eqref{GARCH-parameter}.
Thus, we have
	\begin{equation*}
	\int_{n-1}^{n} \sigma_{t}(\theta)^{2}dt = h_{n}(\theta) + D_n,
	\end{equation*}
where $D_n = D_n^c + D_n^J$. 
Since the integrand of $D_n^c$ is predictable, $D_n$ is a martingale difference.
Proposition \ref{Lemma : integrated vol} (b) and (c) can be showed immediately following the results of Proposition \ref{Lemma : integrated vol} (a).
\endpf

%%%%%%%%%%%%
%%%%%%%%%%%%
%%%%%%%%%%%%

\subsection{Proof of Theorem \ref{Thm-GH-rate}}

Maximizing $\hat{L}^{GH}_{n,m}$ proposed in Section \ref{SEC-3.3} is equivalent to maximizing
	\begin{equation*}
	\hat{L}^{GH}_{n,m} = -\frac{1}{2n} \sum_{i=1}^n \left[ \log (\hat{h}_i(\theta)) + \frac{RV_i}{\hat{h}_i(\theta)} \right].
	\end{equation*}
We focus on $\hat{L}^{GH}_{n,m}$ defined above in this proof.
Define
	\begin{align*}
	\hat{L}_{n,m} ^{GH}( \theta) =& -\frac{1} {2n} \sum _{i=1} ^{n} \left[ \mbox{log} (\hat{h}_i (\theta)) +\frac{RV_i}{\hat{h}_i(\theta)} \right] =-\frac{1}{2n} \sum_{i=1}^{n} \hat{l}_i^{GH}(\theta) \quad \text{and} \quad \hat{\psi}_{n,m}^{GH}(\theta) =\frac{\partial \hat{L}_{n,m}^{GH}(\theta)}{\partial \theta}; \\
	\hat{L}_{n}^{GH}( \theta) =& -\frac{1}{2n} \sum _{i=1} ^{n} \left[ \mbox{log} (h_i (\theta)) +\frac{ \int_{i-1}^i \sigma_t^2 (\theta_0) dt}{h_i(\theta)} \right] \quad \text{and} \quad \hat{\psi}_n^{GH}(\theta)=\frac{\partial \hat{L}_{n}^{GH}(\theta) }{\partial \theta}; \\
	L_{n}^{GH}(\theta) =& -\frac{1} {2n} \sum _{i=1} ^{n} \left[ \mbox{log} (h_i (\theta)) + \frac{h_i (\theta _0)}{h_i(\theta)} \right] \quad \text{and} \quad \psi_n^{GH}(\theta  )=\frac{\partial L_{n}^{GH}(\theta)}{\partial \theta}.
	\end{align*}
To ease notations, we denote derivatives of any given function $g$ at $x_0$ by
	\begin{equation*}
	\frac{\partial g(x_0)}{\partial x} = \frac{\partial g(x)}{\partial x} \Bigg|_{x=x_0}.
	\end{equation*}
Lemma 1 in \citet{kim2016unified} shows that the dependence of $h_i(\theta)$ on the initial value decays exponentially.
Thus, we may use the true initial value $\sigma^2_0(\theta_0)$ during the rest of the proofs.

\begin{lemma}
	\label{Mis}
	Under Assumption \ref{assumption : GH} (a)-(f), we have
	\begin{enumerate}
		\item[(a)]  $E \left( R_i^2 \right)=E \left( \int_{i-1}^{i} \sigma_t^2 (\theta_0)dt \right) =E \left \lbrace h_i(\theta_0) \right \rbrace,$ $\sup_{i \in \mathbb{N}} E(R^2_i) \leq \frac{\omega^g_0 + \beta_0^g \lambda \omega_L}{1-\alpha^g_0-\gamma_0} + E(h_1(\theta_0)) < \infty, $
		and
		$ \sup_{i \in \mathbb{N}} E(\sup_{\theta \in \Theta} h_i(\theta)) < \infty;$ 
		
		\item[(b)] for any $p \geq 1$,
		\begin{eqnarray*}
		&&\sup_{i \in \mathbb{N}} \left\| \sup_{\theta \in \Theta} \hat{h}^{-1}_i (\theta) \frac{\partial \hat{h}_i (\theta)}{\partial \theta_j} \right\|_{L_p} \leq C, \quad \sup_{i \in \mathbb{N}} \left\| \sup_{\theta \in \Theta} \hat{h}^{-1}_i (\theta) \frac{\partial^2 \hat{h}_i (\theta)}{\partial \theta_j \partial \theta_k} \right\|_{L_p} \leq C, \cr
		&& \text{and } \sup_{i \in \mathbb{N}} \left\|  \sup_{\theta \in \Theta} \hat{h}^{-1}_i (\theta) \frac{\partial^3 \hat{h}_i (\theta)}{\partial \theta_j \partial \theta_k \partial \theta_l} \right\|_{L_p} \leq C
		\end{eqnarray*}
		for any $j,k,l \in \lbrace 1,2,3,4 \rbrace$, where $\theta=(\theta_1, \theta_2, \theta_3, \theta_4)=(\omega^g, \alpha^g, \beta^g, \gamma)$.
	\end{enumerate}
\end{lemma}

\textbf{Proof of Lemma \ref{Mis}}. The statements can be showed similar to the proofs of Lemma 2  \citep{kim2016unified}.
\endpf

%%%%%%%%%%%%
%%%%%%%%%%%%
%%%%%%%%%%%%

\begin{lemma}
	\label{GH : bound}
	Under Assumption \ref{assumption : GH} (a)-(d), we have
	\begin{enumerate}
		\item[(a)] there exists a neighborhood $B(\theta_0)$ of $\theta_0$ such that $$\sup_{i \in \mathbb{N}} \left\| \sup _{\theta \in B(\theta_0) } \frac{\partial ^3 \hat{l}_i ^{GH} (\theta)}{\partial \theta_{j}  \partial \theta_{k}  \partial \theta_{l}  } \right\|_{L_1} < \infty$$
		for any  $j,k,l \in \{ 1,2,3,4 \} $ where $\theta =(\theta_1, \theta_2, \theta_3, \theta_4 )=(\omega^g, \alpha^g,   \beta^g, \gamma)$;
		
		\item[(b)] $-\triangledown \psi _n ^{GH} (\theta _0  )$ is a positive definite matrix for $n \geq 5$.
	\end{enumerate}
\end{lemma}
\textbf{Proof of Lemma \ref{GH : bound}}. The proof is in the online Appendix.

%%%%%%%%%%%%
%%%%%%%%%%%%
%%%%%%%%%%%%

\begin{lemma}
	\label{GH : L-consistency}
	Under Assumption \ref{assumption : GH} (a)-(f), we have
	\begin{eqnarray}
	 && \sup_{\theta \in \Theta} \left |\hat{L}_{n,m}^{GH}(\theta)-\hat{L}_n ^{GH}(\theta) \right | = O_p(m^{-1/4}),  \label{Lemma-L-result1} \\
	 && \sup_{\theta \in \Theta}\left | \hat{L}_{n}^{GH}(\theta)-L_n^{GH}(\theta) \right | =o_p(1), \label{Lemma-L-result2} \\ 
	&&\sup_{\theta \in \Theta} \left |\hat{L}_{n,m}^{GH}(\theta)-L_n^{GH}(\theta) \right | = O_p(m^{ -1/4}) +o_p(1).  \label{Lemma-L-result3}
	\end{eqnarray}
\end{lemma}
\textbf{Proof of Lemma \ref{GH : L-consistency}}. The proof is in the online Appendix.

%%%%%%%%%%%%
%%%%%%%%%%%%
%%%%%%%%%%%%

\begin{proposition} \label{GH : consistency}
	Under  Assumption \ref{assumption : GH} (a)-(d), there is a unique maximizer of $L_n^{GH} (\theta)$ and as $m,n \rightarrow \infty$, $\hat{\theta}^{GH} \rightarrow \theta_0  $ in probability.
\end{proposition}

\textbf{Proof of Proposition \ref{GH : consistency}.}
The statement can be showed similar to the proofs of Theorem 1 \citep{kim2016unified} together with the result of Lemma \ref{GH : L-consistency}.
\endpf
\\

%%%%%%%%%%%%
%%%%%%%%%%%%
%%%%%%%%%%%%

\textbf{Proof of Theorem \ref{Thm-GH-rate}.}
By the mean value theorem and Taylor expansion, there exists $ \theta^*$ between $\theta_0 $ and $\hat{\theta}^{GH}$ such that
	\begin{equation*}
	\hat{\psi} _{n,m} ^{GH} (\theta _0 )- \hat{\psi} _{n,m} ^{GH} (\hat{\theta}^{GH}) =  \hat{\psi} _{n,m}^{GH} (\theta _0 )=- \triangledown \hat{\psi} _{n,m}^{GH} ( \theta^*) (\hat{\theta}^{GH} -\theta _0 ).
	\end{equation*}
If $-\triangledown \hat{\psi} _{n,m}^{GH} (\theta^*)\overset{p}{\rightarrow}  -\triangledown \psi _{n} ^{GH} (\theta_0 )$ which is a positive definite matrix by Lemma \ref{GH : bound} (b), the convergence rate of $\|\hat{\theta}^{GH} -\theta _0\|_{max}$ is the same as that of $\hat{\psi} _{n,m}^{GH} (\theta _0 )$.
Thus, it is enough to show
$$\hat{\psi} _{n,m} ^{GH}(\theta _0 )= O_p(m^{-1/4}) +O_p (n^{-1/2})$$
and
$$\quad \left \|\triangledown \hat{\psi} _{n,m} ^{GH} (\theta ^*)  -\triangledown \psi _n ^{GH} (\theta_0 ) \right \|_{max} =o_p(1).$$
First consider $ \hat{\psi} _{n,m} ^{GH}(\theta _0 )= O_p(m^{-1/4}) +O_p (n^{-1/2})$.
Similar to the proofs of Theorem 2 \citep{kim2016unified}, we can show that
	\begin{eqnarray}	\label{eq000-Thm1}
	\hat{\psi} _{n,m}^{GH} (\theta _0)&=&\psi _{n}^{GH} (\theta _0) +\frac{1}{2n}\sum_{i=1}^{n}\frac{\partial h_i(\theta_0)}{\partial \theta } h_i(\theta_0)^{-1} \frac{D_i}{h_i(\theta_0)} +O_p(m^{-1/4})\cr
	&=&\frac{1}{2n} \sum_{i=1}^{n}\frac{\partial h_i(\theta_0)}{\partial \theta } h_i(\theta_0)^{-1} \frac{D_i}{h_i(\theta_0)} +O_p(m^{-1/4}).
	\end{eqnarray}
By the application of the  It\^o's lemma  and It\^o's isometry,  we can show for any $j \in \{1,2,3,4\}$,
	\begin{eqnarray*}	
	%\label{equation-GHpf-2}
	&& E\[ \(\frac{1}{2n} \sum_{i=1}^{n}\frac{\partial h_i(\theta_0)}{\partial \theta _j} h_i(\theta_0)^{-1} \frac{D_i}{h_i(\theta_0)} \)^2\] \cr
	&&=  \frac{1}{4n^2} \sum_{i=1}^{n} E\[ \( \frac{\partial h_i(\theta_0)}{\partial \theta _j} \)^2 h_i(\theta_0)^{-2} \frac{E\[ D_i^2 \middle | \FF_{i-1} \]}{h_i ^2 (\theta_0)}\] \cr
	&&\leq   \frac{C}{ n^2} \sum_{i=1}^{n} E\[ \( \frac{\partial h_i(\theta_0)}{\partial \theta _j} \)^2 h_i(\theta_0)^{-2} \frac{1}{h_i ^2 (\theta_0)}\] \cr
	&&\leq  \frac{C}{n^2} \sum_{i=1}^{n} E\[ \( \frac{\partial h_i(\theta_0)}{\partial \theta _j} \) ^2 h_i(\theta_0)^{-2} \]  \leq  C n^{-1},
	\end{eqnarray*}
where the last inequality is due to Lemma \ref{Mis} (b).
Similar to the proofs of Theorem 2 \citep{kim2016unified} together with the results of Lemma \ref{GH : bound} and Proposition \ref{GH : consistency}, we can show  
	\begin{equation*}
	\left \|\triangledown \hat{\psi} _{n,m} ^{GH} (\theta ^*)  -\triangledown \psi _n ^{GH} (\theta_0) \right \|_{max} =o_p(1).
	\end{equation*}
\endpf

%%%%%%%%%%%%
%%%%%%%%%%%%
%%%%%%%%%%%%

\subsection{Proof of Theorem \ref{Thm-GH-normality}}

\textbf{Proof of Theorem \ref{Thm-GH-normality}.}
By the mean value theorem and  Taylor expansion,  we have for some $ \theta^*$ between $\theta_0 $ and $\hat{\theta}^{GH}$,
	\begin{eqnarray*}
	  - \triangledown \hat{\psi} _{n,m}^{GH} ( \theta^*) (\hat{\theta}^{GH} -\theta _0 )  &=&  \hat{\psi} _{n} ^{GH} (\theta _0)  + \left \{ \hat{\psi} _{n,m}^{GH} (\theta _0 )-\hat{\psi} _{n} ^{GH} (\theta _0) \right \} \cr
	& =& \frac{1}{2n} \sum_{i=1}^{n}\frac{\partial h_i(\theta_0)}{\partial \theta } h_i(\theta_0)^{-1} \frac{D_i}{h_i(\theta_0)}   + O_p ( m^{-1/4}),
	\end{eqnarray*}
where the second equality is due to \eqref{eq000-Thm1}.
By the ergodic theorem and the result in the proof of Theorem \ref{Thm-GH-rate},  we have
$$ -\triangledown \hat{\psi} _{n,m} ^{GH} (\theta ^*) \to  B   \; \text{ in probability}, $$
and $B$ is a positive definite matrix.
For any $f \in \mathbb{R}^4$, let
$$d_i = f ^T \frac{\partial h_i(\theta _0) }{\partial \theta  } h_i(\theta_0) ^{-1} \frac{D_i} {h_i(\theta_0)}.$$
Then $d_i$ is a martingale difference with  $E(d_i^2 ) <\infty$.

Since $\left( D_i, \int_{i-1}^i \sigma_t ^2 (\theta_0) dt, R_i^2 \right)$'s are stationary and ergodic processes, $d_i$ is also stationary and ergodic.
By the martingale central limit theorem and  Cram$\acute{\text{e}}$r-Wold device,  we have
\begin{equation*}
-\sqrt{n}\hat{\psi} _{n} ^{GH} (\theta _0) =\sqrt{n}\frac{1}{2n} \sum_{i=1}^{n} \frac{\partial h_i(\theta _0) }{\partial \theta} h_i(\theta_0) ^{-1} \frac{D_i} {h_i(\theta_0)} \overset{d} {\rightarrow} N(0,A  ^{GH}).
\end{equation*}
Therefore, by Slutsky's theorem, we conclude that
	\begin{eqnarray*}
	\sqrt{n} (\hat{\theta}^{GH} -\theta_0) 	&\overset{d}{\rightarrow}& N(0,  B ^{ -1} A^{GH} B^{ -1}).
	\end{eqnarray*}
\endpf

%%%%%%%%%%%%
%%%%%%%%%%%%
%%%%%%%%%%%%

\subsection{Proof of Theorem \ref{Thm-GHQ-rate2}}

Maximizing $\hat{L}^{GHO}_{n,m}$ is equivalent to maximizing 	
	\begin{equation*}
	\hat{L}^{GHO}_{n,m}(\phi) = - \frac{1}{2n} \sum_{i=1}^n \left[ \log ( \hat{h}_i( \theta ) ) +\frac{RV_i}{\hat{h}_i(\theta)} \right] - \frac{1}{2n} \sum_{i=1}^{n} \left[ \log \sigma_e^2 + \frac{(NV_{i-1} - \hat{f}_i(\varphi) )^2}{\sigma_e^2} \right],
	\end{equation*}
where $\hat{f}_i(\varphi) = b + a \hat{h}_{i}(\theta)$.
We focus on $\hat{L}^{GHO}_{n,m}$ defined above in this proof. 
Define
	\begin{eqnarray*}
	\hat{L}^{GHO}_{n,m}(\phi) &=& - \frac{1}{2n} \sum_{i=1}^n \left[ \log ( \hat{h}_i( \theta ) ) +\frac{RV_i}{\hat{h}_i(\theta)} \right] - \frac{1}{2n} \sum_{i=1}^{n} \left[ \log \sigma_e^2 + \frac{(NV_{i-1} - \hat{f}_i(\varphi) )^2}{\sigma_e^2} \right] \cr
	&=&  - \frac{1}{2n} \sum_{i=1}^n \hat{l}^{GH}_i (\theta)  - \frac{1}{2n} \sum_{i=1}^{n} \hat{l}^{GO}_i (\phi)
	\end{eqnarray*}
and
$$\hat{\psi}_{n,m}^{GHO}(\phi)=\frac{\partial\hat{L}_{n,m}^{GHO}(\phi) }{\partial\phi};$$
	\begin{equation*}
	\hat{L}^{GHO}_{n}(\phi) =- \frac{1}{2n} \sum_{i=1}^n \left[ \log ( h_i( \theta ) ) +\frac{\int_{i-1}^i \sigma^2_t (\theta_0) dt}{h_i(\theta)} \right] - \frac{1}{2n} \sum_{i=1}^{n} \left[ \log \sigma_e^2 + \frac{(NV_{i-1} - f_i(\varphi) )^2}{\sigma_e^2} \right]
	\end{equation*}
and
$$ \hat{\psi}_{n}^{GHO}(\phi)=\frac{\partial\hat{L}_{n}^{GHO}(\phi) }{\partial\phi}; $$
	\begin{equation*}
	L_{n}^{GHO}(\phi)=-\frac{1}{2n}\sum_{i=1}^{n}\left(\log h_{i}(\theta)+\frac{h_{i}(\theta_{0})}{h_{i}(\theta)}\right)-\frac{1}{2n}\sum_{i=1}^{n}\left\{\log\sigma_{e}^{2}+\frac{\left[f_i(\varphi)-f_i(\varphi_0)\right]^{2}+ \sigma_{e0}^{2}}{\sigma_{e}^{2}}\right\}
	\end{equation*}
and
$$\psi_{n}^{GHO}(\phi)=\frac{\partial{L}_{n}^{GHO}(\phi) }{\partial\phi}.$$

%%%%%%%%%%%%
%%%%%%%%%%%%
%%%%%%%%%%%%

\begin{lemma}
	\label{GHQ : bound2}
	Under Assumption \ref{assumption : GH} (a)--(f) and Assumption \ref{assumption-GHO} (a)--(b),  
	\begin{enumerate}
		\item[(a)] there exists a neighborhood $B(\phi_{0})$ around $\phi_0$ such that $$\sup\limits_{i\in \mathbb{N}}\left\|\sup\limits_{\phi\in B(\phi_{0})} \frac{\partial^{3}\hat{l}_{i}^{GO}(\phi)}{\partial\phi_{j} \partial\phi_{k}\partial\phi_{l}}\right\|_{L_{1}}<\infty$$ for any $j,k,l \in \left\{1,2,\ldots, 7 \right\}$, where $\phi=(\phi_{1},\phi_{2}, \phi_{3},\phi_{4},\phi_{5},\phi_{6}, \phi_{7})=(\omega^g,\alpha^g, \beta^g, \gamma ,a,b,\sigma^2_e)$;
		
		\item[(b)] $-\triangledown \psi_{n}^{GHO}(\theta_{0})$ is a  positive definite matrix for $n \geq 7$.
	\end{enumerate}
\end{lemma}

\textbf{Proof of Lemma \ref{GHQ : bound2}.} The proof is in the online Appendix. \endpf

%%%%%%%%%%%%
%%%%%%%%%%%%
%%%%%%%%%%%%

\begin{lemma}
	\label{GHQ : consistent2}
	Under Assumption \ref{assumption : GH} (a)-(f) and Assumption \ref{assumption-GHO} (a)--(b),  we have
	\begin{eqnarray*}
	&& \sup_{\phi \in \Phi } \left|\hat{L}_{n,m}^{GHO}(\phi)-\hat{L}_n^{GHO}(\phi)\right| = O_p(m^{-1/4}), \cr
	&& \sup_{\phi \in \Phi } \left|\hat{L}_{n}^{GHO}(\phi)-L_n^{GHO}(\phi)\right| = o_p(1),\cr
	&& \sup_{\phi \in \Phi } \left|\hat{L}_{n,m}^{GHO} (\phi)-L_n^{GHO}(\phi)\right|= O_p(m^{-1/4})+o_p(1).
	\end{eqnarray*}
\end{lemma}

\textbf{Proof of Lemma \ref{GHQ : consistent2}.} The proof is in the online Appendix. \endpf

%%%%%%%%%%%%
%%%%%%%%%%%%
%%%%%%%%%%%%

\begin{proposition}
	\label{GHQ : maximizer2}
	Under Assumption \ref{assumption : GH} (a)-(f) and Assumption \ref{assumption-GHO} (a)--(b),  there exists a unique maximizer for $L_{n}^{GHO}(\phi)$.
	As $m,n \rightarrow\infty$, $\hat{\phi}^{GHO}\rightarrow \phi_{0}$ in probability, where $\phi_0$ is a vector of true parameters.
\end{proposition}

\textbf{Proof of Proposition \ref{GHQ : maximizer2}.}
According to the definition of $L_{n}^{GHO}(\phi)$, we have
	\begin{align*}
	\max \limits_{\phi\in\Phi}~L_{n}^{GHO}(\phi) \leq
	&-\frac{1}{2n}\sum_{i=1}^{n}\min\limits_{\theta_i\in\Theta }~\left[\log(h_{i}(\theta_i))+\frac{h_{i}(\theta_0)}{h_{i}(\theta_i)}\right]\\
	&-\frac{1}{2n}\sum_{i=1}^{n}\min\limits_{\phi_{i}\in\Phi} ~\left[\log\sigma_{ei}^{2}+\frac{\left(f_i(\varphi_i)-f_i(\varphi_0)\right)^{2}+\sigma_{e0}^{2}}{\sigma_{ei}^{2}}\right].
	\end{align*}
Then, similar to the proofs in Theorem 1 of \cite{kim2016unified}, we can show the uniqueness of the solution of $L_{n}^{GHO}(\phi)$, which together with Lemma \ref{GHQ : consistent2} implies  Proposition \ref{GHQ : maximizer2}.
\endpf
\\

%%%%%%%%%%%%
%%%%%%%%%%%%
%%%%%%%%%%%%

\textbf{Proof of Theorem \ref{Thm-GHQ-rate2}.}
By the mean value theorem and Taylor expansion, we have
	\begin{equation*}
	\hat{\psi}_{n,m}^{GHO}(\hat\phi^{GHO})-\hat{\psi}_{n,m}^{GHO}(\phi_{0})=-\hat{\psi}_{n,m}^{GHO}(\phi_{0})=\nabla\hat{\psi}_{n,m}^{GHO}(\phi^{*})(\hat\phi^{GHO}-\phi_{0}),
	\end{equation*}
where $ \phi^{*}$ is between $\phi_{0}$ and $\tilde\phi^{GHO}$.
According to Lemma \ref{GHQ : bound2} (b), $-\nabla \psi_{n}^{GHO}(\phi_{0})$ is a positive definite matrix.
If $-\nabla \hat{\psi}_{n,m}^{GHO}(\phi^{*}) \xrightarrow{p} -\nabla \psi_{n}^{GHO}(\phi_{0})$, then the convergence rate of $\hat\phi^{GHO} - \phi_{0}$ is the same as the convergence rate of $\hat{\psi}_{n,m}^{GHO}(\phi_{0})$.

By the similar arguments in the proof of Theorem \ref{Thm-GH-rate}, we can show
	\begin{equation*}
	\left\| \hat{\psi}_{n,m}^{GHO}(\phi_{0})-\hat{\psi}_{n}^{GHO}(\phi_{0}) \right\|_{L_1}\leq  C m^{-1/4}.
	\end{equation*}
We have
	\begin{align}
	\hat{\psi}_{n}^{GHO}(\phi_{0})=&\frac{1}{2n}\sum_{i=1}^{n}
	\begin{pmatrix}
	\frac{D_{i}}{h_{i}^2(\theta_{0})}\frac{\partial h_{i}(\theta_{0})}{\partial\varphi}\\
	0
	\end{pmatrix}
	-\frac{1}{2n}\sum_{i=1}^{n}
	\begin{pmatrix}
	\frac{-2e_{i}}{\sigma_{e0}^{2}}\frac{\partial f_{i}(\varphi_{0})}{\partial \varphi}\\
	\frac{1}{\sigma_{e0}^{2}}-\frac{e_{i}^{2}}{\sigma_{e0}^{4}}
	\end{pmatrix}.\label{eq:order_1 new}
	\end{align}
The  arguments in the proof of Theorem \ref{Thm-GH-rate} shows that the first term of the right side of (\ref{eq:order_1 new}) is $O_{p}(n^{-1/2})$.
Since $e_{i}$ is independent of $\frac{\partial f_{i}(\varphi_{0})}{\partial \varphi}$, the second term of the right side of (\ref{eq:order_1 new}) is also $O_{p}(n^{-1/2})$.
Thus, the convergence rate of $\hat{\psi}_{n,m}^{GHO}(\phi_{0})$ is $n^{-1/2} + m^{-1/4}$.

Similar to the proof of Theorem \ref{Thm-GH-rate},  we can show
	\begin{equation*}
	\left\|\nabla \hat{\psi}_{n,m}^{GHO}(\phi^{*})-\nabla\psi_{n}^{GHO}(\phi_{0})\right\|_{\max}=o_p(1).
	\end{equation*}
Therefore, the statement is proved.
\endpf

%%%%%%%%%%%%
%%%%%%%%%%%%
%%%%%%%%%%%%

\subsection{Proof of Theorem \ref{Thm-GHQ-normality2}}

\textbf{Proof of Theorem \ref{Thm-GHQ-normality2}.}
Since   the mean value theorem and Taylor expansion provides
	\begin{equation*}
	\hat{\psi}_{n,m}^{GHO}(\hat\phi^{GHO})-\hat{\psi}_{n,m}^{GHO}(\phi_{0})=-\hat{\psi}_{n,m}^{GHO}(\phi_{0})=\nabla\hat{\psi}_{n,m}^{GHO}(\phi^{*})(\hat\phi^{GHO}-\phi_{0}),
	\end{equation*}
where $ \phi^{*}$ is between $\phi_{0}$ and $\hat\phi^{GHO}$, we have
	\begin{align*}
	\sqrt{n}(\hat\phi^{GHO}-\phi_{0})=&-\sqrt{n}\left(\nabla\psi_{n}^{GHO}(\phi_{0})+o_{p}(1)\right)^{-1}\hat{\psi}_{n}^{GHO}(\phi_{0})+o_{p}(1),
	\end{align*}
where the equality can be showed similar  to the proof of Theorem \ref{Thm-GH-rate}.
Since $e_{i}$ is independent of $D_{i}$ and $\left(D_{i},e_{i},Z_{i}^{2}\right)$ is stationary and ergodic, by the Cram\'{e}r-Wold device and the martingale central limit theorem, we have
	\begin{equation*}
	\sqrt{n}\hat{\psi}_{n}^{GHO}(\phi_{0})=\frac{\sqrt{n}}{2n}\sum_{i=1}^{n}
	\begin{pmatrix}
	\frac{D_{i}}{h_{i}^2(\theta_{0})}\frac{\partial h_{i}(\theta_{0})}{\partial\varphi}\\
	0
	\end{pmatrix}
	-\frac{\sqrt{n}}{2n}\sum_{i=1}^{n}
	\begin{pmatrix}
	\frac{-2e_{i}}{\sigma_{e0}^{2}}\frac{\partial f_{i}(\varphi_{0})}{\partial \varphi}\\
	\frac{1}{\sigma_{e0}^{2}}-\frac{e_{i}^{2}}{\sigma_{e0}^{4}} 
	\end{pmatrix} \overset{d}{\rightarrow} N(0,A^{GHO}).
	\end{equation*}
On the other hand, we have
	\begin{align*}
	&-\nabla \psi_{n}^{GHO}(\phi_{0})\\
	&=\frac{1}{2n}
	\begin{pmatrix}
	\sum_{i=1}^{n}\frac{\partial h_{i}(\theta_{0})}{\partial\varphi}\frac{\partial
	h_{i}(\theta_{0})}{\partial\varphi^{T}}h_{i}(\theta_{0})^{-2}&\mathbf{0} _{6\times1}\\
	\mathbf{0}_{6\times1}^{T}&0\\
	\end{pmatrix}+\frac{1}{2n}
	\begin{pmatrix}
	\frac{2}{\sigma_{e0}^{2}}\sum_{i=1}^{n}\frac{\partial f_{i}(\varphi_{0})}{\partial\varphi}\frac{\partial f_{i}(\varphi_{0})}{\partial\varphi^{T}}&\mathbf{0}_{6\times1}\\
	\mathbf{0}_{6\times1}^{T}&\frac{n}{\sigma_{e0}^{4}}\\
	\end{pmatrix}\\
	&\overset{p}{\rightarrow}    B^{GHO}.
	\end{align*}
Therefore, by the Slutsky's theorem, we have
	\begin{equation*}
	\sqrt{n}(\hat{\phi}^{GHO}-\phi_{0}) \overset{d}{\rightarrow} N(0,(B^{GHO})^{-1}A^{GHO}(B^{GHO})^{-1}).
	\end{equation*}
\endpf

%%%%%%%%%%%%
%%%%%%%%%%%%
%%%%%%%%%%%%

%%%%%%%%%%%%%%%%%%%%%%%%%%%%%%%%%%%%%%%%%%%%%%%%%%%%%%%%%%%%%%%%%%%%%%%%%%%%%%%%%%%%%%%%%%%%%%%%%%%%%%%%%%%%%%%%%%%%%%%%%%%%
 
\vskip 14pt
\noindent {\large\bf Acknowledgements} \\

The research of Xinyu Song was supported by the Fundamental Research Funds for the Central Universities (2018110128), China Scholarship Council (201806485017), and National Natural Science Foundation of China (Grant No. 11871323).
The research of Donggyu Kim was supported in part by KAIST Settlement/Research Subsidies for Newly-hired Faculty grant G04170049 and KAIST Basic Research Funds by Faculty (A0601003029).
The research of Huiling Yuan was supported by the State Scholarship Fund. 
The research of Xiangyu Cui was supported by National Natural Science Foundation of China (71671106). 
The research of Zhiping Lu was supported by Natural Science Foundation of Shanghai (17ZR1409000) and the 111 Project (B14019).
The research of Yong Zhou was supported by the National Natural Science Foundation of China (71931004 and 91546202).
The research of Yazhen Wang was supported in part by NSF Grants DMS-15-28735, DMS-17-07605, and DMS-19-13149.

We thank the Associate Editor, Viktor Todorov, and two anonymous referees for many constructive suggestions that have significantly improved the paper.

This research was performed using the compute resources and assistance of the UW-Madison Center For High Throughput Computing (CHTC) in the Department of Computer Sciences. The CHTC is supported by UW-Madison, the Advanced Computing Initiative, the Wisconsin Alumni Research Foundation, the Wisconsin Institutes for Discovery, and the National Science Foundation, and is an active member of the Open Science Grid, which is supported by the National Science Foundation and the U.S. Department of Energy's Office of Science. 

%%%%%%%%%%%
%%%%%%%%%%%
%%%%%%%%%%%

\bibliography{myReferences}

\begin{thebibliography}{}

\bibitem[Admati and Pfleiderer, 1988]{admati1988theory}
Admati, A.~R. and Pfleiderer, P. (1988).
\newblock A theory of intraday patterns: Volume and price variability.
\newblock {\em The Review of Financial Studies}, 1(1):3--40.

\bibitem[A{\"\i}t-Sahalia et~al., 2010]{ait2010high}
A{\"\i}t-Sahalia, Y., Fan, J., and Xiu, D. (2010).
\newblock High-frequency covariance estimates with noisy and asynchronous
  financial data.
\newblock {\em Journal of the American Statistical Association},
  105(492):1504--1517.

\bibitem[A{\"\i}t-Sahalia et~al., 2012]{ait2012testing}
A{\"\i}t-Sahalia, Y., Jacod, J., and Li, J. (2012).
\newblock Testing for jumps in noisy high frequency data.
\newblock {\em Journal of Econometrics}, 168(2):207--222.

\bibitem[Ait-Sahalia and Yu, 2009]{ait2008high}
Ait-Sahalia, Y. and Yu, J. (2009).
\newblock High frequency market microstructure noise estimates and liquidity
  measures.
\newblock {\em Annals of Applied Statistics}, 3(1):422--457.

\bibitem[Andersen et~al., 2007]{andersen2007roughing}
Andersen, T.~G., Bollerslev, T., and Diebold, F.~X. (2007).
\newblock Roughing it up: Including jump components in the measurement,
  modeling, and forecasting of return volatility.
\newblock {\em The review of economics and statistics}, 89(4):701--720.

\bibitem[Andersen et~al., 2003]{andersen2003modeling}
Andersen, T.~G., Bollerslev, T., Diebold, F.~X., and Labys, P. (2003).
\newblock Modeling and forecasting realized volatility.
\newblock {\em Econometrica}, 71(2):579--625.

\bibitem[Andersen et~al., 1997]{andersen1997intraday}
Andersen, T.~G., Bollerslev, T., et~al. (1997).
\newblock Intraday periodicity and volatility persistence in financial markets.
\newblock {\em Journal of empirical finance}, 4(2-3):115--158.

\bibitem[Andersen et~al., 2019]{andersen2019time}
Andersen, T.~G., Thyrsgaard, M., and Todorov, V. (2019).
\newblock Time-varying periodicity in intraday volatility.
\newblock {\em Journal of the American Statistical Association},
  114(528):1695--1707.

\bibitem[Barndorff-Nielsen et~al., 2008]{barndorff2008designing}
Barndorff-Nielsen, O.~E., Hansen, P.~R., Lunde, A., and Shephard, N. (2008).
\newblock Designing realized kernels to measure the ex post variation of equity
  prices in the presence of noise.
\newblock {\em Econometrica}, 76(6):1481--1536.

\bibitem[Barndorff-Nielsen and Shephard, 2006]{barndorff2006econometrics}
Barndorff-Nielsen, O.~E. and Shephard, N. (2006).
\newblock Econometrics of testing for jumps in financial economics using
  bipower variation.
\newblock {\em Journal of financial Econometrics}, 4(1):1--30.

\bibitem[Black and Scholes, 1973]{black1973pricing}
Black, F. and Scholes, M. (1973).
\newblock The pricing of options and corporate liabilities.
\newblock {\em Journal of political economy}, 81(3):637--654.

\bibitem[Bollerslev, 1986]{bollerslev1986generalized}
Bollerslev, T. (1986).
\newblock Generalized autoregressive conditional heteroskedasticity.
\newblock {\em Journal of econometrics}, 31(3):307--327.

\bibitem[Brockwell and Davis, 2016]{brockwell2016introduction}
Brockwell, P.~J. and Davis, R.~A. (2016).
\newblock {\em Introduction to time series and forecasting}.
\newblock {S}pringer.

\bibitem[Christensen et~al., 2010]{christensen2010pre}
Christensen, K., Kinnebrock, S., and Podolskij, M. (2010).
\newblock Pre-averaging estimators of the ex-post covariance matrix in noisy
  diffusion models with non-synchronous data.
\newblock {\em Journal of Econometrics}, 159(1):116--133.

\bibitem[Corsi et~al., 2010]{corsi2010threshold}
Corsi, F., Pirino, D., and Reno, R. (2010).
\newblock Threshold bipower variation and the impact of jumps on volatility
  forecasting.
\newblock {\em Journal of Econometrics}, 159(2):276--288.

\bibitem[Davies and Tauchen, 2018]{davies2018data}
Davies, R. and Tauchen, G. (2018).
\newblock Data-driven jump detection thresholds for application in jump
  regressions.
\newblock {\em Econometrics}, 6(2):16.

\bibitem[Engle, 1982]{engle1982autoregressive}
Engle, R.~F. (1982).
\newblock Autoregressive conditional heteroscedasticity with estimates of the
  variance of united kingdom inflation.
\newblock {\em Econometrica}, 50(4):987--1007.

\bibitem[Engle and Gallo, 2006]{engle2006multiple}
Engle, R.~F. and Gallo, G.~M. (2006).
\newblock A multiple indicators model for volatility using intra-daily data.
\newblock {\em Journal of Econometrics}, 131(1--2):3--27.

\bibitem[Fan and Kim, 2018]{fan2018robust}
Fan, J. and Kim, D. (2018).
\newblock Robust high-dimensional volatility matrix estimation for
  high-frequency factor model.
\newblock {\em Journal of the American Statistical Association},
  113(523):1268--1283.

\bibitem[Fan and Wang, 2007]{fan2007multi}
Fan, J. and Wang, Y. (2007).
\newblock Multi-scale jump and volatility analysis for high-frequency financial
  data.
\newblock {\em Journal of the American Statistical Association},
  102(480):1349--1362.

\bibitem[Hansen et~al., 2012]{hansen2012realized}
Hansen, P.~R., Huang, Z., and Shek, H.~H. (2012).
\newblock Realized garch: a joint model for returns and realized measures of
  volatility.
\newblock {\em Journal of Applied Econometrics}, 27(6):877--906.

\bibitem[Hong and Wang, 2000]{hong2000trading}
Hong, H. and Wang, J. (2000).
\newblock Trading and returns under periodic market closures.
\newblock {\em The Journal of Finance}, 55(1):297--354.

\bibitem[Jacod et~al., 2009]{jacod2009microstructure}
Jacod, J., Li, Y., Mykland, P.~A., Podolskij, M., and Vetter, M. (2009).
\newblock Microstructure noise in the continuous case: the pre-averaging
  approach.
\newblock {\em Stochastic processes and their applications}, 119(7):2249--2276.

\bibitem[Kim and Wang, 2016]{kim2016unified}
Kim, D. and Wang, Y. (2016).
\newblock Unified discrete-time and continuous-time models and statistical
  inferences for merged low-frequency and high-frequency financial data.
\newblock {\em Journal of Econometrics}, 194(2):220--230.

\bibitem[Kim et~al., 2016]{kim2016asymptotic}
Kim, D., Wang, Y., and Zou, J. (2016).
\newblock Asymptotic theory for large volatility matrix estimation based on
  high-frequency financial data.
\newblock {\em Stochastic Processes and their Applications},
  126(11):3527--3577.

\bibitem[Kl{\"u}ppelberg et~al., 2004]{kluppelberg2004continuous}
Kl{\"u}ppelberg, C., Lindner, A., and Maller, R. (2004).
\newblock A continuous-time garch process driven by a l{\'e}vy process:
  stationarity and second-order behaviour.
\newblock {\em Journal of Applied Probability}, 41(3):601--622.

\bibitem[Mancini, 2004]{mancini2004estimation}
Mancini, C. (2004).
\newblock Estimation of the characteristics of the jumps of a general
  poisson-diffusion model.
\newblock {\em Scandinavian Actuarial Journal}, 2004(1):42--52.

\bibitem[Shephard and Sheppard, 2010]{shephard2010realising}
Shephard, N. and Sheppard, K. (2010).
\newblock Realising the future: forecasting with high-frequency-based
  volatility (heavy) models.
\newblock {\em Journal of Applied Econometrics}, 25(2):197--231.

\bibitem[Todorov, 2009]{todorov2009estimation}
Todorov, V. (2009).
\newblock Estimation of continuous-time stochastic volatility models with jumps
  using high-frequency data.
\newblock {\em Journal of Econometrics}, 148(2):131--148.

\bibitem[Todorov, 2011]{todorov2011econometric}
Todorov, V. (2011).
\newblock Econometric analysis of jump-driven stochastic volatility models.
\newblock {\em Journal of Econometrics}, 160(1):12--21.

\bibitem[Todorov, 2019]{todorov2019nonparametric}
Todorov, V. (2019).
\newblock Nonparametric spot volatility from options.
\newblock {\em The Annals of Applied Probability}, 29(6):3590--3636.

\bibitem[Wang, 2002]{wang2002asymptotic}
Wang, Y. (2002).
\newblock Asymptotic nonequivalence of garch models and diffusions.
\newblock {\em The Annals of Statistics}, 30(3):754--783.

\bibitem[Xiu, 2010]{xiu2010quasi}
Xiu, D. (2010).
\newblock Quasi-maximum likelihood estimation of volatility with high frequency
  data.
\newblock {\em Journal of Econometrics}, 159(1):235--250.

\bibitem[Zhang, 2006]{zhang2006efficient}
Zhang, L. (2006).
\newblock Efficient estimation of stochastic volatility using noisy
  observations: A multi-scale approach.
\newblock {\em Bernoulli}, 12(6):1019--1043.

\bibitem[Zhang, 2011]{zhang2011estimating}
Zhang, L. (2011).
\newblock Estimating covariation: Epps effect, microstructure noise.
\newblock {\em Journal of Econometrics}, 160(1):33--47.

\bibitem[Zhang et~al., 2005]{zhang2005tale}
Zhang, L., Mykland, P.~A., and A{\"\i}t-Sahalia, Y. (2005).
\newblock A tale of two time scales: Determining integrated volatility with
  noisy high-frequency data.
\newblock {\em Journal of the American Statistical Association},
  100(472):1394--1411.

\bibitem[Zhang et~al., 2016]{zhang2016jump}
Zhang, X., Kim, D., and Wang, Y. (2016).
\newblock Jump variation estimation with noisy high frequency financial data
  via wavelets.
\newblock {\em Econometrics}, 4(3):34.

\end{thebibliography}

\end{document}